\documentclass[prb,twocolumn,showpacs,unsortedaddress]{revtex4-1}
\usepackage[margin=0.6in]{geometry}
\usepackage{hyperref}
\usepackage{mathtools}
\usepackage{bm}
\hypersetup{
  colorlinks   = true, 
  urlcolor     = blue, 
  linkcolor    = blue, 
  citecolor   = red 
}
\usepackage{graphicx}
\usepackage{amsmath}
\renewcommand{\vec}{\mathbf}

\begin{document}

\title{The Intrinsic Magnetization of Antiferromagnetic Textures}

\author{Erlend G. Tveten$^{1}$, Tristan M\"{u}ller$^{2}$, Jacob Linder$^{1}$ and Arne Brataas$^{1}$}
\affiliation{$^1$Department of Physics, Norwegian University of Science and Technology, NO-7491 Trondheim, Norway}
\affiliation{$^2$JARA-Institute for Quantum Information, RWTH Aachen University, D-52074 Aachen, Germany}

\date{\today}

\begin{abstract}
Antiferromagnets (AFMs) exhibit intrinsic magnetization when the order parameter spatially varies. This intrinsic spin is present even at equilibrium and can be interpreted as a twisting of the homogeneous AFM into a state with a finite spin. Because magnetic moments couple directly to external magnetic fields, the intrinsic magnetization can alter the dynamics of antiferromagnetic textures under such influence. Starting from the discrete Heisenberg model, we derive the continuum limit of the free energy of AFMs in the exchange approximation and explicitly rederive that the spatial variation of the antiferromagnetic order parameter is associated with an intrinsic magnetization density. We calculate the magnetization profile of a domain wall and discuss how the intrinsic magnetization reacts to external forces. We show conclusively, both analytically and numerically, that a spatially inhomogeneous magnetic field can move and control the position of domain walls in AFMs. By comparing our model to a commonly used alternative parametrization procedure for the continuum fields, we show that the physical interpretations of these fields depend critically on the choice of parametrization procedure for the discrete-to-continuous transition. This can explain why a significant amount of recent studies of the dynamics of AFMs, including effective models that describe the motion of antiferromagnetic domain walls, have neglected the intrinsic spin of the textured order parameter.
\end{abstract}


\maketitle
\section{Introduction}
Measuring the ordered state of antiferromagnets (AFMs) is complicated by the absence of macroscopic magnetization. The promise of AFMs as candidates for active roles in spintronics logic elements have increased the interest in addressing this problem.\cite{MacDonald13082011,GomonayLTP2013} In particular, the observation of tunneling anisotropic magnetoresistance in AFMs~\cite{Park:2011rt,PhysRevLett.108.017201,Marti:2014xy,PhysRevX.4.041034} represents a clear experimental procedure to detect the antiferromagnetic order. Furthermore, current-induced torques on the antiferromagnetic order have been theoretically predicted~\cite{PhysRevB.73.214426,PhysRevB.83.054428,PhysRevB.84.094404,PhysRevB.81.144427} and experimentally indicated in spin valve systems.~\cite{PhysRevLett.98.116603} Also the ferromagnetic concept of spin pumping has been generalized to AFMs.~\cite{PhysRevLett.113.057601} The possibility of manipulating the antiferromagnetic order parameter by external forces has fueled renewed theoretical interest in domain wall motion in AFMs due to both charge~\cite{PhysRevLett.106.107206, JPhysCondMat.24.024223, PhysRevLett.110.127208} and spin~\cite{PhysRevB.89.081105, tvetenPRL14, PhysRevB.90.104406} currents. However, the reports on current-induced domain wall motion\cite{PhysRevB.79.134423} are based on indirect observations and not confirmed by other methods or groups. Therefore, there is no straightforward method to reliably detect the dynamics of textures in the antiferromagnetic order.

In this article, we discuss the intrinsic magnetization associated with an inhomogeneous antiferromagnetic order parameter. We describe the origin of the intrinsic spin and discuss whether it can be exploited to detect antiferromagnetic texture dynamics, e.g., domain wall motion. To revisit this topic, which was pioneered for one-dimensional systems in Refs.~\onlinecite{PhysRevB.51.15062, PhysRevLett.74.1859, 1995FizNT..21..355I}, we construct the continuum free energy functional for AFMs from the discrete Heisenberg Hamiltonian in the exchange approximation. We use the Hamiltonian approach to show that the intrinsic magnetization due to textures in the order parameter arises from a parity-breaking term in the energy functional that is absent in a commonly used alternative parametrization of the continuum fields. We clarify the mapping between the two different parametrizations and explain how the intrinsic magnetization can be easily missed in models which are based on the alternative parametrization. We further describe the shape of the intrinsic magnetization density for an antiferromagnetic domain wall and discuss its physical significance as a twisting of the homogeneous spinless AFM into a state with a finite spin. The intrinsic magnetization adds up in two- and three-dimensional extended domain wall systems and can affect the dynamics of domain walls subject to external magnetic fields and spin-polarized currents. We discuss how these consequences can go beyond that of the purely quantum topological effects\cite{Braun:2005fk} observed in one-dimensional spin chains.

Studies of domains in AFMs and descriptions of the shape and properties of antiferromagnetic domain walls date back several decades.\cite{bar1979nonlinear,Baryakhtar:1980qf,PhysRevLett.50.1153,SovPhysUspBaryakhtar1985} However, most of the experimental evidence of such domains were restricted to studies of AFMs in which the collinearity of the sublattices is broken due to Dzyaloshinskii-Moriya (DM) anisotropy. In these studies, when the DM field or the external field vanishes so does the equilibrium magnetization of the AFM. Consequently, the detection of domain walls and their dynamics in compensated AFMs remains an experimental challenge. However, antiferromagnetic domain walls are known to exist and have been experimentally observed, e.g., in monolayers of antiferromagnetic Fe,\cite{Bode:2006uq} in the elemental AFM Cr,\cite{PhysRevLett.98.117206} and in the antiferromagnetic insulator NiO.\cite{PhysRevLett.91.237205} Antiferromagnetic domains and domain walls can also be tailored by manipulating the ferrimagnetic precursor layer before cooling the AFM below the N\'{e}el temperature.\cite{PhysRevLett.106.107201} Observation of individual domains in AFMs can be done, e.g., using X-ray magnetic linear dichroism.\cite{PhysRevB.73.020401, doi:10.1021/nl1025908}

A key aspect of detecting the \textit{dynamics} of antiferromagnetic domain walls is whether such solitons of staggered magnetic order are associated with a spatially constricted magnetization density. Ref.~\onlinecite{PhysRevB.51.15062} argued that such a magnetization exists and that the earlier studies of antiferromagnetic spin chains missed certain parity-breaking terms in the transition from the discrete spin model to the continuum approximation. The antiferromagnetic Heisenberg Hamiltonian has been mapped to the non-linear $\sigma$ model for the continuous staggered order parameter.\cite{bar1979nonlinear,Baryakhtar:1980qf,PhysRevLett.50.1153} However, in Haldane's seminal work on large-spin Heisenberg AFMs,\cite{PhysRevLett.50.1153} no apparent parity-breaking terms survived the transition to the continuum model. In \textit{Haldane's mapping}\cite{CabraPujol2004,auerbach2012interacting} the continuum field that is conjugate to the antiferromagnetic order parameter describes the dynamic magnetization only (see Sec.~\ref{subsec:Haldanemapping}). Using a slightly different parametrization of the antiferromagnetic order and the magnetization field, Ivanov \textit{et al.}\cite{PhysRevLett.74.1859,1995FizNT..21..355I}~later demonstrated that the energy functional based on the one-dimensional antiferromagnetic Heisenberg model indeed contains a parity-breaking term in the continuum limit and that this term must be taken into account to describe the equilibrium magnetization of a domain wall. The parity-breaking term included in  Refs.~\onlinecite{PhysRevLett.74.1859,1995FizNT..21..355I} is not equivalent to the well-known "topological $\Theta$ terms"\cite{AffleckNPB1985,Haldane:1985sf}, which arise in effective $\sigma$ model Lagrangians for one-dimensional antiferromagnetic spin chains and are responsible for quantum effects such as Haldane's conjecture.\cite{PhysRevLett.50.1153,Haldane:1985sf,CabraPujol2004}

The recently increased interest in AFMs as active spintronics components has spawned a number of effective models for antiferromagnetic dynamics.\cite{PhysRevLett.106.107206,PhysRevLett.110.127208,PhysRevB.89.081105,tvetenPRL14,PhysRevB.90.104406} These recent models mostly adopt the non-linear $\sigma$ model without introducing a Hamiltonian that includes parity-breaking terms that lead to the intrinsic magnetization of antiferromagnetic textures. The absence of parity-breaking terms in these models may be due to different definitions of the continuum fields, or these terms may have been disregarded in the transition to the continuum limit due to specialized symmetry requirements, which only hold for homogeneous AFMs. Whether the intrinsic magnetization of extended two- and three-dimensional systems can lead to qualitatively new physics for the dynamics of antiferromagnetic textures under the influence of external forces remains an open question that we seek to address in this article. 

The intrinsic magnetization of antiferromagnetic textures is small. A domain wall in a one-dimensional antiferromagnetic spin chain exhibits intrinsic magnetization that is in total no larger than the spin of one sublattice~\cite{PhysRevB.51.15062,PhysRevLett.74.1859}. It is therefore unlikely that such a small magnetic moment can be directly detected in the near future. However, the presence of the small spin of domain walls in one-dimensional spin chains manifests itself through quantum effects.~\cite{PhysRevB.56.8886,Braun:2005fk} In higher-dimensional extended systems, such as synthetic AFMs, the magnetization of a textured multilayer may be of appreciable size~\cite{PapanicolaouJPCM1998}. Furthermore, in thin films or in bulk AFMs, which is the focus of our study, the intrinsic magnetization of a transverse domain wall is additive in the perpendicular directions. The result is a macroscopic magnetization that can be more easily excited and detected and that can influence the dynamics of AFMs beyond that of purely quantum effects.

The paper is organized as follows. In the next section (\ref{sec:Theory}), we take the continuum limit of the Heisenberg Hamiltonian, describe the origin of the intrinsic magnetization, and discuss the consequences for the antiferromagnetic dynamic equations. We also compare our model to Haldane's alternative mapping of the continuum fields. This comparison demonstrates that the continuum fields in these two parametrization procedures have critically different physical interpretations. In Sec.~\ref{seq:DWMotion}, we describe the magnetization profile of a domain wall and discuss generalizations to higher-dimensionsal systems. We show how the intrinsic magnetization leads to qualitatively new physics and that the domain wall can be moved by a spatially inhomogeneous magnetic field that couples to the intrinsic magnetization. In Sec.~\ref{sec:Numerics}, we present numerical results for the motion and control of an antiferromagnetic domain wall and show that we can create potential wells for the domain wall with spatially constricted magnetic fields. In Sec.~\ref{sec:Discussion}, we discuss the experimental consequences of the intrinsic magnetization for extended systems in 2D and 3D. Sec.~\ref{sec:Conclusion} concludes the discussion.

\section{Theory}
\label{sec:Theory}
Our starting point is the Heisenberg Hamiltonian due to the exchange coupling between classical spin vectors on a lattice~\cite{PhysRev.86.694}

\begin{equation}
H=J\sum_{\langle\boldsymbol{\alpha},\boldsymbol{\beta}\rangle}\vec{S}_{\boldsymbol{\alpha}}\cdot\vec{S}_{\boldsymbol{\beta}}\,,
\label{eq:HeisenbergHamiltonian}
\end{equation}
where the positive exchange energy, $J>0$, describes an antiferromagnetic ground state. $\langle\boldsymbol{\alpha},\boldsymbol{\beta}\rangle$ denotes a sum over all nearest neighbor lattice sites described by the two sublattices $\boldsymbol{\alpha}$ and $\boldsymbol{\beta}$, where each spin at $\boldsymbol{\alpha}$ has $N_{n}$ nearest neighbors of type $\boldsymbol{\beta}$, and vice versa. $\boldsymbol{\alpha}$ and $\boldsymbol{\beta}$ are $D$-dimensional vectors, where $D$ is the dimensionality of the AFM. We proceed by describing the simplest model, the $D=1$ antiferromagnetic linear spin chain with easy-axis anisotropy, and later generalize our results to 2D and 3D in Appendix \ref{app:2d}. The focus of our subsequent sections are on extended 3D AFMs in which the order parameter varies along one dimension only.
 
\subsection{Free energy functional for 1D}
\label{sec:Theory1}
We consider a linear spin chain with $2N$ atomic lattice sites, where the spins on half of the lattice sites, denoted by $\alpha$, minimize their energy by pointing in the opposite direction of the spins on their $N_{n}=2$ nearest-neighbor lattice sites, denoted by $\beta$, and vice versa. For the AFM, we impose the boundary conditions that the spin on the left end of the spin chain is of type $\alpha$, whereas the right end of the chain is occupied by a $\beta$ site. Therefore, in the ground state, the AFM is fully compensated, and the total spin vanishes. We define the $z$ axis as the magnetic easy axis. The classical Heisenberg Hamiltonian including the easy-axis anisotropy is
\begin{equation}
H_{\rm{1D}}=J\sum_{\langle\alpha,\beta\rangle}^{N,N}\vec{S}_{\alpha}\cdot\vec{S}_{\beta}-K\left(\sum_{\alpha}^{N}S_{\alpha z}^2+\sum_{\beta}^{N}S_{\beta z}^2\right)\,,
\label{eq:HeisenbergHamiltonian1d}
\end{equation}
where $K$ is the anisotropy energy. In typical easy-axis AFMs, the exchange energy dominates, $|J|\gg |K|$. The classical ground state of the Hamiltonian (\ref{eq:HeisenbergHamiltonian1d}) is degenerate, $(\vec{S}_{\alpha},\vec{S}_{\beta})_{0}\to\pm(S \hat{z}, - S\hat{z})$, where $S$ (in units of $\hbar$) is the spin on a single atomic lattice site.

We now introduce the standard definitions (see Sec.~\ref{subsec:Haldanemapping} for a comparison with an alternative definition that is occasionally mistaken to be equivalent to the present model) of the magnetic and staggered order parameters, $\vec{m}_{i}$ and $\vec{l}_{i}$, on a two-sublattice linear lattice parametrized by $i$: 
\begin{subequations}
\label{eqs:nmdefinition}
\begin{eqnarray}
\vec{m}_{i}&=&\frac{\vec{S}_{\alpha}^{i}+\vec{S}_{\beta}^{i}}{2S}\,,
\label{eq:mparameter}\\
\vec{l}_{i}&=&\frac{\vec{S}_{\alpha}^{i}-\vec{S}_{\beta}^{i}}{2S}\,,
\label{eq:nparameter}
\end{eqnarray}
\end{subequations}
where we have paired the sublattice spins $\vec{S}_{\alpha}^{i}$ and $\vec{S}_{\beta}^{i}$ at unit cell $i$ running over a total of $N$ antiferromagnetic unit cells. In this convention, $\vec{m}_{i}^{2}+\vec{l}_{i}^{2}=1$ and the spins in unit cell $i$ can be expressed as follows:
\begin{subequations}
\label{eqs:sdefinition}
\begin{eqnarray}
\vec{S}_{\alpha}^{i}&=&S(\vec{m}_{i}+\vec{l}_{i})\,,
\label{eq:salpha}\\
\vec{S}_{\beta}^{i}&=&S(\vec{m}_{i}-\vec{l}_{i})\,.
\label{eq:sbeta}
\end{eqnarray}
\end{subequations}

After introducing the magnetization vector $\vec{m}_{i}$ and the staggered order parameter $\vec{l}_{i}$, the Heisenberg Hamiltonian (\ref{eq:HeisenbergHamiltonian1d}) reduces to a sum over antiferromagnetic lattice points:
\begin{eqnarray}
H_{\rm{1D}} & = & J S^{2}\sum_{i}^{N-1}(\vec{m}_{i}-\vec{l}_{i})[(\vec{m}_{i}+\vec{l}_{i})+(\vec{m}_{i+1}+\vec{l}_{i+1})]\nonumber\\
& & +JS^{2}(\vec{m}_{N}^{2}-\vec{l}_{N}^{2})\nonumber\\
& &-K S^{2}\sum_{i}^{N}\left[(\vec{m}_{i}+\vec{l}_{i})_{z}^{2}+(\vec{m}_{i}-\vec{l}_{i})_{z}^{2}\right]\,.
\label{eq:Heisenberg1dRewritten}
\end{eqnarray}

We continue by using the identities $2\vec{m}_{i}\vec{m}_{i+1}=\vec{m}_{i}^2+\vec{m}_{i+1}^2-(\vec{m}_{i+1}-\vec{m}_{i})^2$ and $(\vec{l}_{i}\vec{m}_{i+1}-\vec{m}_{i}\vec{l}_{i+1})=\vec{l}_{i}(\vec{m}_{i+1}-\vec{m}_{i})-\vec{m}_{i}(\vec{l}_{i+1}-\vec{l}_{i})$ to rewrite the bulk part of Eq.~(\ref{eq:Heisenberg1dRewritten}) as follows:
\begin{eqnarray}
H_{\rm{1D}} &\approx & 2J S^{2}\sum_{i}^{N}(\vec{m}_{i}^2-\vec{l}_{i}^2)\nonumber\\
& &+\frac{J S^{2}}{2}\sum_{i}^{N-1}\left[(\vec{l}_{i+1}-\vec{l}_{i})^2-(\vec{m}_{i+1}-\vec{m}_{i})^2\right]\nonumber\\
& &+ JS^2\sum_{i}^{N-1}\left[\vec{m}_{i}(\vec{l}_{i+1}-\vec{l}_{i})-\vec{l}_{i}(\vec{m}_{i+1}-\vec{m}_{i})\right]\nonumber\\
& &-2K S^{2}\sum_{i}^{N}(m_{i,z}^2+l_{i,z}^{2})\,,
\label{eq:Heisenberg1dRewritten2}
\end{eqnarray} 
where we have disregarded the vanishingly small energy contribution $-JS^{2}(\vec{m}_{1}^{2}+\vec{m}_{N}^{2}-\vec{n}_{1}^{2}-\vec{n}_{N}^{2})/2$ from the unit cells at the edges.

Next we go to the large-$N$ limit and take the continuum approximation, allowing us to write $H_{\rm{1D}}\approx\int(d_{i}/\Delta)\mathcal{H}(\vec{l},\vec{l}',\vec{m},\vec{m}')$, where $\Delta$ is the length of the antiferromagnetic unit cell and $\vec{l}'$ and $\vec{m}'$ are the (dimensionless) spatial derivatives of the staggered field and the magnetization, respectively. $d_{i}$ is an infinitesimal length element along the spin chain. For $D=1$, $\Delta=2d$, where $d$ is the nearest-neighbor spacing in the linear chain. The energy density (apart from a constant and in units of energy) is
\begin{eqnarray}
\mathcal{H}_{\rm{1D}}(\vec{l},\vec{l}',\vec{m},\vec{m}') = J S^{2}[4|\vec{m}|^2+|\vec{l}'| ^2-|\vec{m}'| ^2\nonumber\\
+(\vec{m}\cdot\vec{l}'-\vec{l}\cdot\vec{m}')]-K S^{2}[(\vec{l}\cdot\hat{z})^{2}+(\vec{m}\cdot\hat{z})^{2}]\,.
\label{eq:EnergyDensity}
\end{eqnarray}
We note that the fourth exchange term in Eq.~(\ref{eq:EnergyDensity}) has an unusual parity-breaking form\cite{AffleckNPB1985,0953-8984-1-19-001} and is an odd function of the order parameter $\vec{l}$. 

\begin{figure}
\centering
\includegraphics[scale=0.5]{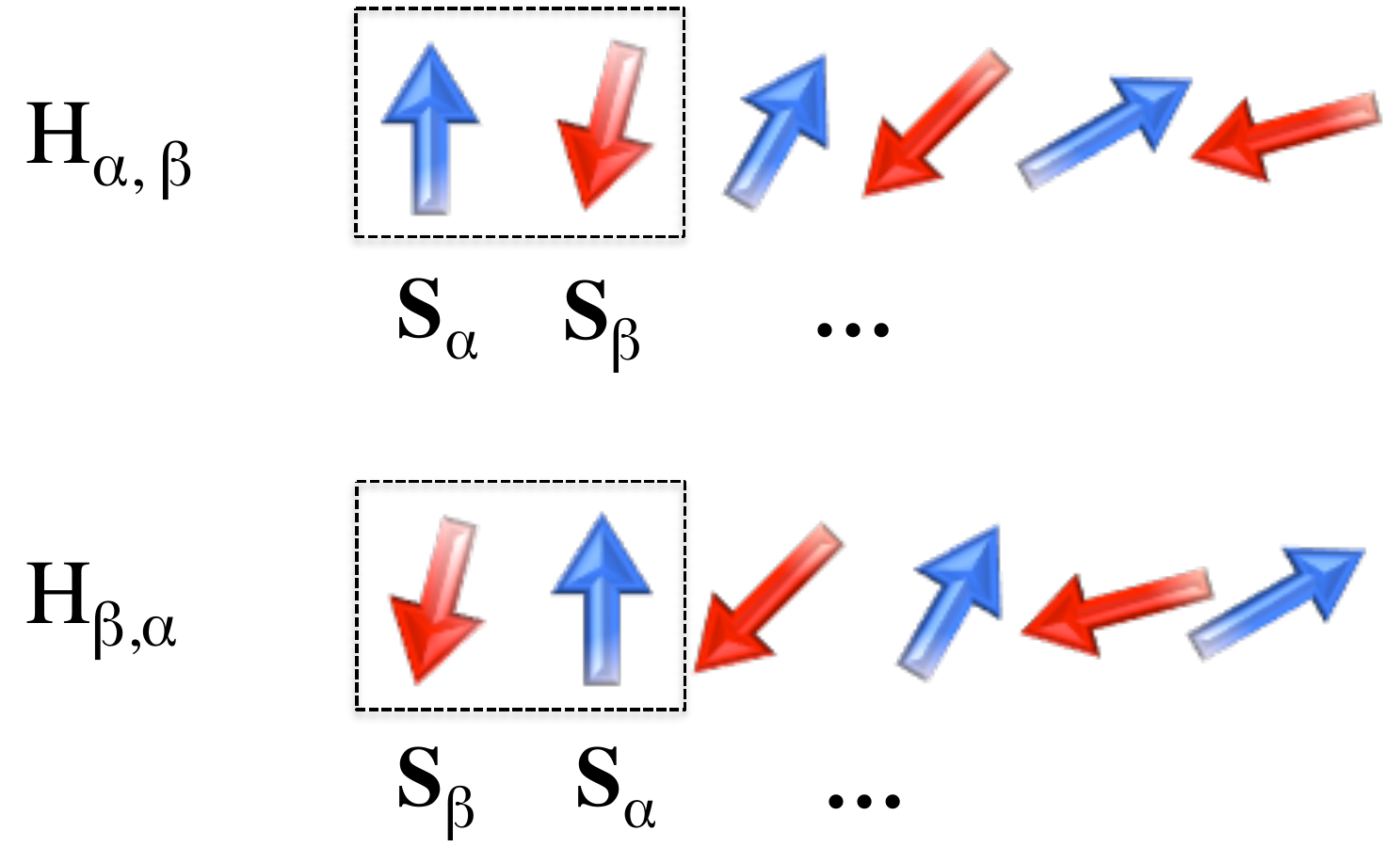}
\caption{For a simple linear spin chain with antiferromagnetic exchange coupling, the Heisenberg Hamiltonian is not invariant under sublattice exchange, $\vec{S}_{\alpha}\leftrightarrow\vec{S}_\beta$, if the order parameter is spatially inhomogeneous. 
(\textbf{Top}) A simplified sketch of a 6-spin 90$^{\circ}$ texture. Exchanging the spins on sublattices $\alpha$ and $\beta$ (\textbf{bottom}) creates a more disordered phase that costs additional exchange energy, hence $H_{\alpha,\beta}<H_{\beta,\alpha}$. In the continuum limit, this energy difference is captured by the parity-breaking term in the antiferromagnetic energy functional.}
\label{fig:sublattice}
\end{figure}

In the models of AFMs that we consider, the two-sublattice linear lattice in 1D, the centered squared lattice in 2D, and the body-centered cubic lattice in 3D, the Heisenberg Hamiltonian is not invariant under sublattice exchange ($\alpha\leftrightarrow\beta$) if the order parameter is spatially inhomogeneous, see Fig.~\ref{fig:sublattice}. However, there is an ambiguity in the pairing of spins $\vec{S}_{\alpha}^{i}$ and $\vec{S}_{\beta}^{i}$ and the definition of the order parameter $\vec{l}_{i}$ in Eq.~(\ref{eq:nparameter}). One might as well choose $\tilde{\vec{l}}_{i}=-\vec{l}_{i}$ as the order parameter, and consequently, one usually demands that the bulk Hamiltonian is invariant under the transformations $\vec{l}_{i}\rightarrow-\vec{l}_{i}$ and $\vec{m}_{i}\rightarrow\vec{m}_{i}$\cite{lifshitz1980} because the two possible choices of the order parameter are physically equivalent. Under these transformations, the definitions of $\vec{S}_{\alpha}^{i}$ and $\vec{S}_{\beta}^{i}$ in Eqs.~(\ref{eqs:sdefinition}) also change, and the fourth exchange term in Eq.~(\ref{eq:EnergyDensity}) undergoes an additional sign change. The energy functional is therefore invariant with respect to the two equivalent definitions of the order parameter but not invariant under sublattice exchange. In the latter case, the ordering of the spins changes, resulting in a larger exchange energy penalty for inhomogeneous AFMs. A simplified sketch of this energy difference is shown in Fig.~\ref{fig:sublattice} for a 6-spin chain with a 90$^{\circ}$ texture. In the bottom spin chain the $\alpha$ and $\beta$ sublattices have been exchanged, leading to a more disordered phase that costs additional exchange energy. This result generalizes to an arbitrary number of spins in a linear textured spin chain.

To describe the order parameter dynamics of the AFM, it is useful to work in the exchange approximation~\cite{lifshitz1980}, $|J|\gg|K|$, and consider slowly varying antiferromagnetic textures. In this case, $|\vec{m}|^{2}\ll |\vec{l}|^{2}$, and we can disregard terms that are of higher order than $|\vec{m}|^{2}$, such as the magnetic anisotropy energy term and the magnetic stiffness term in Eq.~(\ref{eq:EnergyDensity}). We choose the spin chain axis to be along the $z$ axis and introduce the normalized staggered vector field $\vec{n}(z,t)\equiv \vec{l}(z,t)/|\vec{l}(z,t)|$. We can consequently write the energy density as a function of the deviations $\partial_{z}\vec{n}$ ($\equiv\partial\vec{n}/\partial z$) and $\vec{m}$ from the ground state. After integrating by parts, we arrive at the free energy density for the linear antiferromagnetic spin chain to the lowest order in deviations from an equilibrium state:~\cite{1995FizNT..21..355I}
\begin{equation}
\mathcal{H}_{\rm{1D}}(\vec{n},\partial_{z}\vec{n},\vec{m}) = \frac{a}{2}|\vec{m}| ^2+\frac{A}{2}|\partial_{z}\vec{n}| ^2+ L(\vec{m}\cdot\partial_{z}\vec{n})-\frac{K_{z}}{2}(\vec{n}\cdot\hat{z})^{2}\,.
\label{eq:EnergyDensityExchApprox}
\end{equation}
The equation has the following parameters: the homogeneous exchange energy $a=8 JS^{2}$, the exchange stiffness terms $A=\Delta^{2}JS^{2}$ and $L=2\Delta JS^{2}$, and the anisotropy energy $K_{z}=2KS^{2}$. Here, a finite $L$ lifts the degeneracy of the sublattice exchange.

\subsection{Free energy functional for $D>1$}
In Appendix \ref{app:2d}, we generalize the free energy of Eq.~(\ref{eq:EnergyDensityExchApprox}) to 2D and 3D for the centered squared and the body centered cubic unit cell, respectively. We find that the generalized free energy density in the exchange approximation is given by
\begin{eqnarray}
\mathcal{H}(\vec{n},\partial_{i}\vec{n},\vec{m}) &=& \frac{a}{2}|\vec{m}| ^2+\frac{A}{2}\left[\sum_{i}|\partial_{i}\vec{n}| ^2+\frac{1}{2}\sum_{i\neq j}(\partial_{i}\vec{n}\cdot\partial_{j}\vec{n})\right]\nonumber\\
& &+ L\sum_{i}(\vec{m}\cdot\partial_{i}\vec{n})-\frac{K_{z}}{2}(\vec{n}\cdot\hat{z})^{2}\,,
\label{eq:EnergyDensityExchApprox3D}
\end{eqnarray}
where $a=4N_{D} J S^{2}$, $A=N_{D}\Delta^{2} J S^{2}/2$, $L=N_{D}\Delta J S^{2}$, $K_{z}=2 K S^{2}$, and $N_{D}$ is the number of nearest neighbors. $N_{1}=2$, $N_{2}=4$ for the squared lattice, and $N_{3}$ depends on the choice of unit cell, 6 for the simple cubic cell and 8 for the body-centered cubic cell. The stiffness part of the above Hamiltonian density contains two apparent anisotropic terms, $\sim(\partial_{i}\vec{n}\cdot\partial_{j}\vec{n})$ and $\sim(\vec{m}\cdot\partial_{i}\vec{n})$. However, in the following, we show that after eliminating the degrees of freedom associated with $\vec{m}$, the effective Lagrangian reduces to the non-linear $\sigma$ model and the resulting antiferromagnetic spin-wave dispersion remains isotropic . 

This isotropic dispersion is in contrast to the anisotropic dispersion relation resulting from the exchange term identified by Lifshitz and Pitaevskii\cite{lifshitz1980}, which is similar but not identical to the third term in Eq.~(\ref{eq:EnergyDensityExchApprox3D}). Lifshitz and Pitaevskii consider only the small deviation $\vec{n}_{\perp}$ ($\vec{n}\rightarrow\vec{n}_{0}+\vec{n}_\perp$) from the equilibrium homogeneous antiferromagnetic spin configuration and add the exchange term $\sim(\vec{m}\cdot\partial_{z}\vec{n}_{\perp}-\vec{n}_{\perp}\cdot\partial_{z}\vec{m})$ to the free energy density. Compared to Eq.~(\ref{eq:EnergyDensity}), this also results in a surface anisotropy $\sim (\vec{n}_{0}\cdot\partial_{z}\vec{m})$, which (after integration over the space) favors magnetization build-up on the edges of the AFM. Consequently, the dispersion relation for this model is anisotropic. The parity-breaking exchange term ($\sim L$) in the above free energy density (\ref{eq:EnergyDensity}) differs from the term of Lifshitz and Pitaevskii because it involves $\vec{n}$ rather than $\vec{n}_{\perp}$ and does not violate the isotropic dispersion relation of antiferromagnetic spin waves due to small variations in the staggered field $\vec{n}$. This is also the case for $D>1$. Neglecting the parity-breaking term as being of leading order in the exchange energy would imply that an AFM at equilibrium exhibits no intrinsic magnetization, even when textures in the staggered field are present.

\subsection{Lagrangian density and equations of motion}
The equations of motion for the staggered field $\vec{n}$ and the magnetization field $\vec{m}$ can be found from, e.g.,  linear combinations of the equations of motion for the sublattice spins $\vec{S}_{\boldsymbol{\alpha}}$ and $\vec{S}_{\boldsymbol{\beta}}$.\cite{PhysRevB.51.15062} Equivalently, we may proceed by constructing the Lagrangian density and directly compute the dynamic equations for $\vec{n}$ and $\vec{m}$ from the variation of the Lagrangian with respect to these fields. Our starting point is the generalized free energy density in the exchange approximation, Eq.~(\ref{eq:EnergyDensityExchApprox3D}). The Lagrangian density can be constructed as $\mathcal{L}=\mathcal{K}-\mathcal{H}$, where $\mathcal{K}$ is the kinetic energy term. Analogous to the procedure for constructing the kinetic term for a single spin in a ferromagnet,\cite{PhysRevB.53.3237,Tatara2008213} $\mathcal{K}$ can be constructed from the Berry phase of the spin \textit{pair} $\vec{S}_{\boldsymbol{\alpha}}+\vec{S}_{\boldsymbol{\beta}}$ that constitutes the antiferromagnetic unit cell:
\begin{equation}
\label{eq:berryphasepair}
\int{\mathcal{K}dV} =-S\hbar\left[\sum_{\boldsymbol{\alpha}}\vec{A}_{\boldsymbol{\alpha}}\cdot\dot{\vec{S}}_{\boldsymbol{\alpha}}+
\sum_{\boldsymbol{\beta}}\vec{A}_{\boldsymbol{\beta}}\cdot\dot{\vec{S}}_{\boldsymbol{\beta}}\right]\,,
\end{equation}
where it is convenient to choose the gauge potential $\vec{A}_{\boldsymbol{\alpha}(\boldsymbol{\beta})}$ such that the spin-pair Berry phase vanishes in the strictly antiparallel configuration, $\vec{S}_{\boldsymbol{\alpha}}=-\vec{S}_{\boldsymbol{\beta}}$. One such choice is $\vec{A}_{\boldsymbol{\alpha}(\boldsymbol{\beta})}=-\hat{\phi}_{\boldsymbol{\alpha}(\boldsymbol{\beta})}\cos{\theta_{\boldsymbol{\alpha}(\boldsymbol{\beta})}}/\sin{\theta_{\boldsymbol{\alpha}(\boldsymbol{\beta})}}$ in the spherical coordinate system, where $\theta$ is the polar angle and $\hat{\phi}$ is a unit vector along the azimuth. This gauge is identical to that which is normally used to describe the kinetic energy of a single spin in ferromagnets\cite{Tatara2008213} and generalized to a two-sublattice model with antiparallel spin configuration. By expanding the spin pair Berry phase in small deviations from the antiparallel configuration, $\theta_{\boldsymbol{\beta}}\to\pi-(\theta_{\boldsymbol{\alpha}}+\delta\theta)$ and $\phi_{\boldsymbol{\beta}}\to\pi+(\phi_{\boldsymbol{\alpha}}+\delta\phi)$, and transferring back to the $[\vec{n},\vec{m}]$ basis, the kinetic term in the continuum approximation is given by\cite{PhysRevLett.50.1153,PhysRevLett.74.1859}
\begin{equation}
\label{eq:berryphase}
\mathcal{K}=\rho \vec{m}(\dot{\vec{n}}\times\vec{n})\,,
\end{equation}
where $\rho=2S\hbar$ is the magnitude of the staggered spin angular momentum per unit cell and we have disregarded terms of the order $|\vec{m}|^{4}$ and higher.

Varying the Lagrangian with respect to the magnetization $\vec{m}$ and the staggered field $\vec{n}$ gives the coupled Landau-Lifshitz equations of motion
\begin{subequations}
\label{eqs:LLG}
\begin{eqnarray}
\dot{\vec{n}} & = & \vec{\omega}_{\vec{m}}\times\vec{n}\,,
\label{eq:LLGn}\\
\dot{\vec{m}} & = & \vec{\omega}_{\vec{n}}\times\vec{n}+\vec{\omega}_{\vec{m}}\times\vec{m}\,,
\label{eq:LLGm}
\end{eqnarray}
\end{subequations}
where damping is typically phenomenologically introduced.\cite{PhysRevLett.106.107206} In the transverse basis, where $|\vec{n}|^{2}=1$, no term of the form $\sim(\vec{\omega}_{\vec{n}}\times\vec{m})$ (as present in the dynamics of $\vec{l}$ in, e.g.,  Ref.~\onlinecite{PhysRevB.81.144427}) appears in Eq.~(\ref{eq:LLGn}), which is valid in the exchange approximation and includes terms up to second order in small deviations from equilibrium. The effective magnetic and staggered fields (in units of $s^{-1}$) are defined as functional derivatives of the total free energy $H$:
\begin{subequations}
\label{eqs:effectivefields}
\begin{eqnarray}
\rho\vec{\omega}_{\vec{m}}&\equiv -\frac{\delta H}{\delta \vec{m}}= &-a\vec{m}- L\partial^{i}\vec{n}\,,\label{eq:fmdefinition}\\
\rho\vec{\omega}_{\vec{n}}&\equiv -\frac{\delta H}{\delta\vec{n}}=&A(\nabla^{2}\vec{n}+\frac{1}{2}\partial^{i}\partial^{j}\vec{n})\nonumber\\
& &+ L\partial^{i}\vec{m}+K_{z}(\vec{n}\cdot\hat{z})\hat{z}\,,\label{eq:fndefinition}
\end{eqnarray}
\end{subequations}
where we have defined the sum over spatial derivatives in all directions as $\partial^{i}\equiv\sum_{i=x,y,z}\partial/\partial i$ and $\partial^{i}\partial^{j}\equiv\sum_{i\neq j}\partial^{2}/(\partial i\partial j)$. In the Appendix (\ref{app:2d}) we discuss how these anisotropic differential operators arise in 2D and 3D.

In the absence of external forces in the effective magnetic field, Eqs.~(\ref{eq:LLGn}) and (\ref{eq:fmdefinition}) give\cite{PhysRevLett.74.1859} 
\begin{equation}
\label{eq:meq}
\vec{m}=\frac{\rho}{a}\dot{\vec{n}}\times\vec{n}-\frac{L}{a}\partial^{i}\vec{n}\,,
\end{equation}
which indicates that the magnetization field $\vec{m}$ is simply a slave variable that follows the temporal \textit{and} spatial evolution of the staggered field $\vec{n}$. We note that if we neglect the parity-breaking term in the free energy ($L\to 0$), the intrinsic magnetization of a textured AFM vanishes at equilibrium. Our analysis shows that for our particular parametrization of the continuum fields, this parity-breaking term is an important part of the transition from the discrete spin model to the continuum approximation and cannot be disregarded.

Equation (\ref{eq:meq}) allows us to eliminate $\vec{m}$ and write an effective Lagrangian density for the staggered field and its derivatives as
\begin{eqnarray}
\label{eq:efflagrange}
\mathcal{L}(\vec{n},\dot{\vec{n}},\partial_{i}\vec{n})&=&\frac{\rho^{2}}{2a} |\dot{\vec{n}}| ^{2}-\frac{A-L^{2}/a}{2}\sum_{i}|\partial_{i}\vec{n}| ^{2}\nonumber\\
& & +\frac{\rho L}{a}\sum_{i}\partial_{i}\vec{n}\cdot(\vec{n}\times\dot{\vec{n}})+\frac{K_{z}}{2}(\vec{n}\cdot\hat{z})^{2}\,.
\end{eqnarray}
This Lagrangian density describes the anisotropic non-linear $\sigma$ model with a kinetic topological term (third term).\cite{PhysRevLett.50.1153,PhysRevB.38.7215,PhysRevLett.61.1029,0953-8984-1-19-001,Read1989609,PhysRevB.42.4568} This topological term is a by-product of the elimination of $\vec{m}$ from the Lagrangian. It can be shown that this term has the form of a total derivative\cite{0953-8984-1-19-001}. Consequently, it has no effect on the effective equations of motion for $\vec{n}$ or the domain wall dynamics that we describe in the next sections. We will not discuss in any detail the quantum effects of the topological term in the following.

\subsection{Comparison with Haldane's mapping}
\label{subsec:Haldanemapping}
We digress for a moment to compare the one-dimensional model described above with a commonly used  alternative definition of the continuum fields known as Haldane's mapping\cite{PhysRevLett.50.1153,auerbach2012interacting,CabraPujol2004} of the antiferromagnetic order parameter. We include this comparison because the different parametrizations are not equivalent and are, therefore, recurrent sources for confusion. In contrast to the Hamiltonian approach described by Eqs.~(\ref{eqs:nmdefinition}) and (\ref{eqs:sdefinition}), Haldane's parametrization maps each spin in the spin chain at cite $i$ onto two continuum fields:
\begin{equation}
\label{eq:haldanesmapping}
\vec{S}_{i}/S=(-1)^{i}\tilde{\vec{n}}_{i}\sqrt{1-\tilde{\vec{m}}_{i}^{2}}+\tilde{\vec{m}}_{i}\,, 
\end{equation}
where $\tilde{\vec{n}}$ is the unitary N\'{e}el field and $\tilde{\vec{m}}$ is the "canting" field. We note that this mapping introduces extra degrees of freedom, which must subsequently be reduced by limiting the Fourier components of the fields $\tilde{\vec{n}}$ and $\tilde{\vec{m}}$ to include only long-wavelength excitations.\cite{auerbach2012interacting}

\begin{figure}
\centering
\includegraphics[scale=0.45]{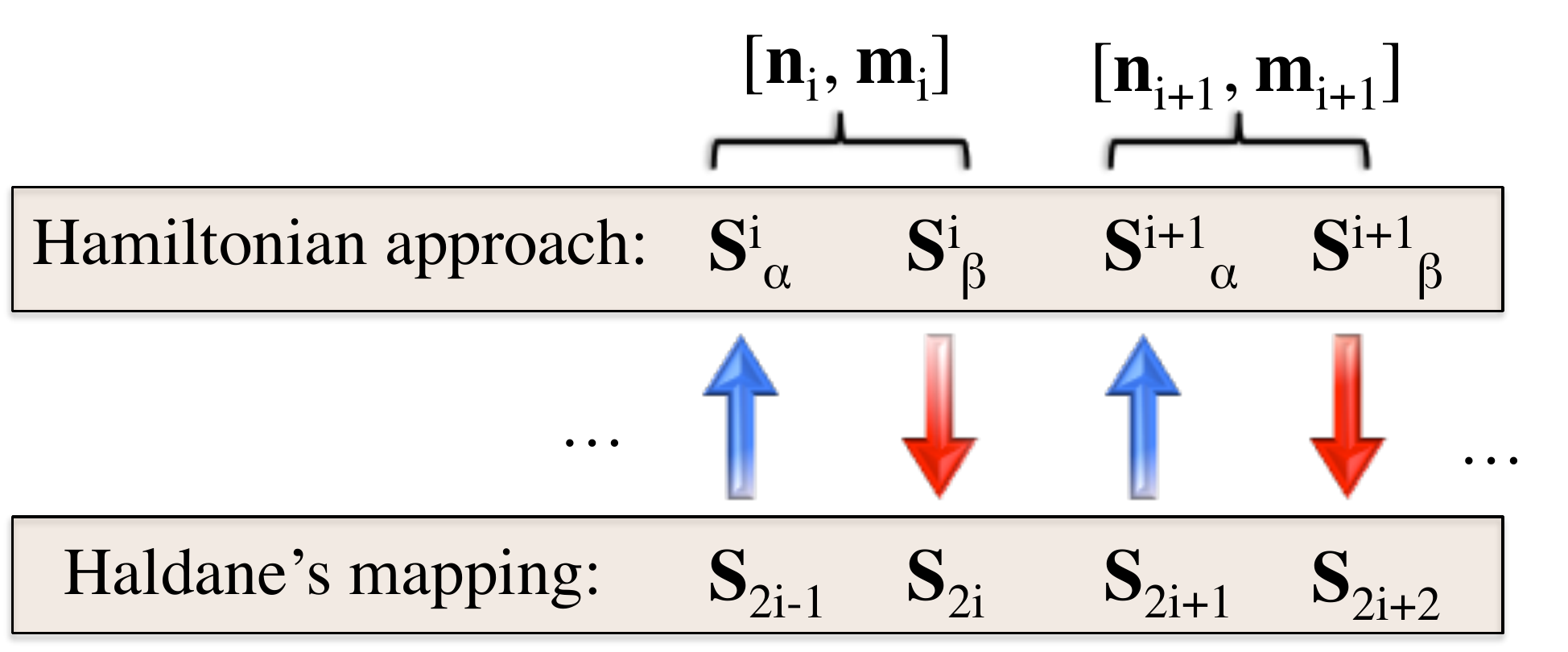}
\caption{In the Hamiltonian approach (\textbf{top}), Eqs.~(\ref{eqs:nmdefinition}) define values for the staggered field $\vec{n}_{i}$ and the magnetization field $\vec{m}_{i}$ at the center of every antiferromagnetic unit cell labelled by $i$. In Haldane's mapping (\textbf{bottom}), every single spin is mapped onto two continuum fields, the N\'{e}el field $\tilde{\vec{n}}$ and the "canting" field $\tilde{\vec{m}}$.}
\label{fig:haldanecomp}
\end{figure}

Figure \ref{fig:haldanecomp} compares the labelling of the spins in the Hamiltonian approach used in this work with that of Haldane's mapping. By equating the expressions for $\vec{S}_{\alpha}^{i}$ and $\vec{S}^{i}_{\beta}$ in Eqs.~(\ref{eqs:sdefinition}) and their corresponding expressions in Haldane's parametrization, we find the relationship between the continuum fields in the two different parametrizations:
\begin{subequations}
\label{eqs:mapping}
\begin{eqnarray}
\vec{m}_{i}+\vec{n}_{i}\sqrt{1-\vec{m}_{i}^{2}}&=&-\tilde{\vec{n}}_{2i-1}\sqrt{1-\tilde{\vec{m}}_{2i-1}^{2}}+\tilde{\vec{m}}_{2i-1}\\
\vec{m}_{i}-\vec{n}_{i}\sqrt{1-\vec{m}_{i}^{2}}&=&\tilde{\vec{n}}_{2i}\sqrt{1-\tilde{\vec{m}}_{2i}^{2}}+\tilde{\vec{m}}_{2i}
\end{eqnarray}
\end{subequations}
In the exchange approximation, $\vec{m}\ll \vec{n}$ and $\tilde{\vec{m}}\ll \tilde{\vec{n}}$. Keeping only the lowest order contributions in the magnetization $\vec{m}$ and the canting field $\tilde{\vec{m}}$, it follows that
\begin{subequations}
\label{eqs:nmapping}
\begin{eqnarray}
\vec{n}_{i} & \approx & -\frac{1}{2}(\tilde{\vec{n}}_{2i-1}+\tilde{\vec{n}}_{2i})+\frac{1}{2}(\tilde{\vec{m}}_{2i-1}-\tilde{\vec{m}}_{2i})\, ,\\
\vec{m}_{i} & \approx & -\frac{1}{2}(\tilde{\vec{n}}_{2i-1}-\tilde{\vec{n}}_{2i})+\frac{1}{2}(\tilde{\vec{m}}_{2i-1}+\tilde{\vec{m}}_{2i})\, ,
\end{eqnarray}
\end{subequations}
where we have disregarded terms of the order $|\vec{m}|^{2}$ and $|\tilde{\vec{m}}|^{2}$ and higher. 

For small-angle spatial variations in the continuum fields, we use the gradient approximation to find the field values for $\tilde{\vec{n}}$ and $\tilde{\vec{m}}$ at the center of each unit cell: $\tilde{\vec{n}}_{i+1/2}\approx \tilde{\vec{n}}_{i}+(\Delta/4)\partial_{z}\tilde{\vec{n}}_{i}$ and $\tilde{\vec{m}}_{i+1/2}\approx \tilde{\vec{m}}_{i}+(\Delta/4)\partial_{z}\tilde{\vec{m}}_{i}$, where $\Delta/2=d$ is the nearest neighbor distance and $\tilde{\vec{n}}(\tilde{\vec{m}})_{i+1/2}$ represents the N\'{e}el (canting) field at the midpoint between the spins $\vec{S}_{i}$ and $\vec{S}_{i+1}$. Inserting these lowest order gradient approximations into Eqs.~(\ref{eqs:nmapping}) results in a one-to-one relationship between the continuum fields of the Hamiltonian approach and Haldane's parametrization. Correspondingly, the mapping between the two different representations reduces to $\vec{n}\to-\tilde{\vec{n}}+(\Delta/4)\partial_{z}\tilde{\vec{m}}+\mathcal{O}(|\tilde{\vec{m}}|^{2})$ and $\vec{m}\to\tilde{\vec{m}}-(\Delta/4)\partial_{z}\vec{\tilde{n}}+\mathcal{O}(|\tilde{\vec{m}}|^{2})$.

It is critically important that the continuum fields $\tilde{\vec{n}}$ and $\tilde{\vec{m}}$ of Haldane's mapping are not identical to the staggered and magnetization fields $\vec{n}$ and $\vec{m}$ used in the present work. By inserting the mapping between the two parametrizations into the energy functional in Eq.~(\ref{eq:EnergyDensityExchApprox}) and keeping only terms of the order $|\tilde{\vec{m}}|^{2}$ in the exchange approximation, we find the continuum limit energy functional of Haldane's mapping:
\begin{equation}
\mathcal{H}_{\rm{Hal}}(\tilde{\vec{n}},\partial_{z}\tilde{\vec{n}},\tilde{\vec{m}}) = \frac{a}{2}|\tilde{\vec{m}}|^2+\frac{A}{2}|\partial_{z}\tilde{\vec{n}}|^2-\frac{K_{z}}{2}(\tilde{\vec{n}}\cdot\hat{z})^{2}\,.
\label{eq:EnergyDensityExchApproxHaldane}
\end{equation}
This result conclusively shows that the parity-breaking exchange term in Eq.~(\ref{eq:EnergyDensityExchApprox}), which is a result of the procedure of breaking the lattice into spin pairs, vanishes after a suitable transformation of the continuum fields, e.g., $\vec{m}\to\tilde{\vec{m}}-(\Delta/4)\partial_{z}\tilde{\vec{n}}$. In other words, when applying Haldane's mapping procedure, the parity-breaking exchange term does not appear in the energy functional. An overall requirement, however, is that the physics remains the same, including the existence of the intrinsic magnetization.

Although the Hamiltonian approach used in the present work and Haldane's mapping are both valid continuum representations of spin systems with antiferromagnetic exchange coupling, a crucial difference exists for the physical interpretations of the continuum fields, which are \textit{not} equivalent in the two representations. The equilibrium value of the canting field $\tilde{\vec{m}}$ of Haldane's mapping vanishes, also when $\tilde{\vec{n}}$ is inhomogeneous. Therefore, $\tilde{\vec{m}}$ represents the dynamic magnetization induced by temporal variations of the order parameter $\tilde{\vec{n}}$ and not the total magnetization. Consequently, the coupled equations of motion for $\tilde{\vec{n}}$ and $\tilde{\vec{m}}$ are not of the same form as Eqs.~(\ref{eqs:LLG}) and (\ref{eqs:effectivefields}). In particular, the expression for the canting field: $\tilde{\vec{m}}\sim\dot{\tilde{\vec{n}}}\times{\tilde{\vec{n}}}$, which is analogous to Eq.~(\ref{eq:meq}), does not include a term proportional to the gradient of $\tilde{\vec{n}}$. This fact may be an important reason why the intrinsic magnetization is easily missed in models based on Haldane's mapping.

In the Hamiltonian approach, on the other hand, $\vec{m}$ can be interpreted as a magnetization density in the sense that the total accumulated spin (both intrinsic and dynamical) of the AFM can be found from integration, $\vec{M}/S=\int \vec{m}\ dV$. For antiferromagnetic textures, this integral is generally nonzero even for static spin systems. Although the canting field $\tilde{\vec{m}}$ in Haldane's mapping does not include the intrinsic contribution to the magnetization density, the total spin can instead be found from the relation $\vec{M}/S\approx\sum_{i=1}^{2N}[(-1)^{i}\tilde{\vec{n}}(z_{i})+\tilde{\vec{m}}(z_{i})]$. The intrinsic magnetization can be identified as arising from the first terms in the sum. For a slowly varying $\tilde{\vec{n}}$ in, e.g., the $\hat{z}$ direction, $\sum_{i=1}^{2N}(-1)^{i}\tilde{\vec{n}}(z_{i})\cdot\hat{z}\approx [\tilde{n}_{z}(z_{1})-\tilde{n}_{z}(z_{2N})]/2$,\cite{PhysRevB.38.7215} which is generally nonzero for textured AFMs. In the following analysis, we continue using the Hamiltonian approach, in which the continuum field $\vec{m}$ is interpreted as the total magnetization density.

\subsection{Antiferromagnetic spin waves and spin current}
\label{sec:DispersionRelation}
To study small harmonic excitations from a homogeneous AFM, we construct the effective equation of motion for the staggered vector field $\vec{n}$ by combining Eqs.~(\ref{eq:LLGn}) and (\ref{eq:LLGm}) while retaining the constraint $|\vec{n}|^{2}=1$:
\begin{equation}
\label{eq:nlinresponse}
\vec{n}\times(\ddot{\vec{n}}\times\vec{n})=\frac{1}{\rho^{2}}\vec{n}\times\left[(aA-L^{2})\nabla^{2}\vec{n}+aK_{z}(\vec{n}\cdot\hat{z})\hat{z}\right]\times\vec{n}\,.
\end{equation}
The parity-breaking exchange term leads to the renormalization of the exchange stiffness $A\to A^{*}=(A-L^{2}/a)=A/2$ but otherwise leaves the equation of motion (\ref{eq:nlinresponse}) invariant in linear response.\cite{SovPhysUsp23.21} The topological term in Eq.~(\ref{eq:efflagrange}) has no effect on the effective equations of motion for $\vec{n}$, as expected.

Insertion of a small harmonic excitation from the ground state in time and space, $\vec{n}(\vec{r},t)\rightarrow \hat{z}+\delta\vec{n}_{\perp}\exp{[i(\omega t - \vec{k}\cdot\vec{r})]}$ into Eq.~(\ref{eq:nlinresponse}) results in the usual "relativistic" antiferromagnetic dispersion relation
\begin{equation}
\label{eq:dispersion}
\omega^{2}=\frac{1}{\rho^{2}}[aA^{*}k^{2}+aK_{z}]\,,
\end{equation}
where $k=|\vec{k}|$. In the isotropic limit, $K_z\to 0$, which results in the familiar linear dispersion
\begin{equation}
\label{eq:dispersionisotropic}
\omega_{i} =ck\,,
\end{equation}
where $c=N_{n}SJ\Delta/(2\hbar)$ is the spin wave phase velocity. For $\Delta=2d/\sqrt{D}$, where $d$ is the nearest-neighbor distance, and for hypercubic lattices, where $N_{n}=2D$, Eq.~(\ref{eq:dispersionisotropic}) agrees with Eqs.~(13) and (20) in the semi-classical treatment in Ref.~\onlinecite{PhysRev.86.694}, as well as with Holstein-Primakoff calculations\cite{PhysRev.117.117,PhysRevB.40.2494} and Haldane's $D=1$ result.\cite{PhysRevLett.50.1153} We note that the parity-breaking term ($\sim L$) does not lead to an anisotropic dispersion relation, such as the term in Lifshitz and Pitaevskii.\cite{lifshitz1980} On the contrary, the inclusion of such a term is important to arrive at the correct dispersion relation in the classical continuum limit.

The intrinsic magnetic moment of antiferromagnetic textures will necessarily influence how spin currents in inhomogeneous AFMs are described. A continuity equation for the spin angular momentum transfer in the AFM caused by the exchange interaction can be constructed from Eq.~(\ref{eq:LLGm}) as $\dot{\rho\vec{m}}+\sum_{i}\partial_{i}\vec{J}_{s,i}=0$. The spin current polarized along $i$ is
\begin{equation}
\vec{J}_{s,i}=A^{*}\partial_{i}\vec{n}\times\vec{n}-\frac{\rho L}{a}\dot{\vec{n}}\,,
\label{eq:SpinCurrent}
\end{equation}
where we have used Eq.~(\ref{eq:meq}) to eliminate $\vec{m}$. Equation (\ref{eq:SpinCurrent}) explicitly shows that a time-varying antiferromagnetic texture is equivalent to spin angular momentum transfer, a relationship that easily can be missed by models for the staggered dynamics that disregard the intrinsic magnetization. This result may have implications for antiferromagnetic spin pumping from textures\cite{PhysRevLett.113.057601} because the collective motion of the antiferromagnetic order parameter is equivalent to a current of spin angular momentum. In one-dimensional textures, $\rho L/a= S\hbar d$, thus indicating that textures that oscillate at frequency $T^{-1}$ produce a spin-current corresponding approximately to a single spin moving one lattice spacing per period of oscillation $T$.

\subsection{Consequences for staggered dynamics}
In effective models for the dynamics of the staggered vector field $\vec{n}$, the magnetization field $\vec{m}$ plays the role of a slave variable that follows the temporal and spatial evolution of $\vec{n}$. When no external forces couple directly to the intrinsic spin in the AFM, the parity-breaking term in the energy functional ($\sim L$) only leads to a renormalization of the exchange stiffness and has no other effect on the dynamic equations. However, we show in the following that by including external magnetic fields or spin-polarized currents, the dynamics of the antiferromagnetic order parameter can also be altered indirectly through the excitation of the magnetization density field $\vec{m}$.

The spin-transfer torque on \textit{ferromagnetic} textures is normally considered a second-order effect in AFMs when acting only on the small magnetization $\vec{m}(t)$ induced by the time variation of the staggered field, $\dot{\vec{n}}$. If AFMs also exhibit intrinsic magnetization, the spin-transfer torque from spin-polarized currents on the magnetization $\vec{m}$ may become more important. However, because the magnetization is first order in the spatial variation of the staggered field, $\vec{m}\sim\partial_{i}\vec{n}$, the Berger spin transfer torques (Eqs.~(5) and (6) in Ref.~\onlinecite{Brataas:2012fk}) are of the order $\sqrt{K/J}$ smaller than the driving forces acting directly on textures in the staggered field, first identified in Ref.~\onlinecite{PhysRevB.89.081105}. In this case, the intrinsic magnetization of AFMs will lead to higher-order corrections to the current-induced torques that couple directly to the staggered field. In antiferromagnetic thin films with strong surface anisotropy or in special cases in which a strained geometry suppresses the torques on the staggered field, the Berger torques on the textured magnetization could become important. We will not discuss the effects of spin-polarized currents any further in the following. 

Instead, we focus on the effect of an external magnetic field $\vec{H}$ that couples directly to the intrinsic magnetization of antiferromagnetic textures. To illustrate this phenomenon, we add the Zeeman interaction to the free energy density, $\mathcal{H}_{\vec{H}}=\mathcal{H}-\rho\gamma(\vec{H}\cdot\vec{m})$, where $\gamma$ is the gyromagnetic ratio. The external magnetic field induces a small magnetic moment density in the AFM, and the magnetization field $\vec{m}$ is altered according to
\begin{equation}
\label{eq:meqHfield}
\vec{m}=\frac{\rho}{a}\dot{\vec{n}}\times\vec{n}-\frac{L}{a}\partial^{i}\vec{n}+\frac{\gamma\rho}{a}\vec{n}\times(\vec{H}\times\vec{n})\,,
\end{equation}
where the cross products enforce the constraint $\vec{n}\cdot\vec{m}=0$. Inserting this result in the Lagrangian gives the effective Lagrangian density for an AFM under the influence of an external magnetic field $\vec{H}$:
\begin{eqnarray}
\label{eq:EffLagrangianH}
\mathcal{L}_{\vec{H}} & = & \frac{\rho^{2}}{2a}(\dot{\vec{n}}-\gamma\vec{H}\times\vec{n})^{2}-\frac{A^{*}}{2}\sum_{i}(\partial_{i}\vec{n})^{2}\nonumber\\
& &+\frac{K_{z}}{2}(\vec{n}\cdot\hat{z})^{2} + \frac{\rho L}{a}\sum_{i}\partial_{i}\vec{n}\cdot(\vec{n}\times\dot{\vec{n}})\nonumber\\
& &-\frac{\gamma\rho L}{a}\sum_{i}\vec{H}\cdot\partial_{i}\vec{n}\,.
\end{eqnarray}
This Lagrangian density agrees with that proposed in Ref.~\onlinecite{SovPhysUsp23.21}, with the exception of the second to last topological term and the last term, which couples the external magnetic field and  textures in the antiferromagnetic order. In the following, we show how this coupling between magnetic fields and the gradient of the staggered field allows the movement of domain walls in AFMs to be controlled by spatially varying magnetic fields. This result has not been reported previously.

Utilizing the method of collective coordinates\cite{PhysRevLett.100.127204,PhysRevLett.110.127208}, we assume that the temporal dependence of the staggered vector field $\vec{n}(\vec{r},t)$ is held by a set of coordinates $\{a_{j}(t)\}$ that describe the time evolution of textures in the AFM, such that $\vec{n}(\vec{r},\{a_{j}(t)\})$. In this case, the time derivative of the staggered field can be written as $\dot{\vec{n}}=\sum_{j}\dot{a}_{j}\partial_{a_{j}}\vec{n}$. We earlier demonstrated that in AFMs, the collective coordinates can be viewed as quasi-particles with an effective mass reacting to external forces and following Newton's 2nd law.\cite{PhysRevLett.110.127208} The equation of motion for the collective mode $a_{j}$ is
\begin{equation}
\label{eq:CollectiveCoordinatesEquation}
M^{ij}(\ddot{a}_{j}+\frac{a\alpha}{\rho}\dot{a}_{j})=F^{i}\,,
\end{equation}
where $M^{ij}=(\rho^{2}/a)\int\rm{dV}(\partial_{a_{i}}\vec{n}\cdot\partial_{a_{j}}\vec{n})$ is the effective mass, $\alpha$ is the phenomenological Gilbert damping parameter for AFMs, and $F^{i}$ are the forces that act on the collective excitations. $F^{i}=F_{\rm{int}}^{i}+F_{\rm{ext}}^{i}$ can be split into the internal exchange forces $F_{\rm{int}}^{i}=\partial_{a_{i}}H$, which are derivatives of the free energy with respect to the collective modes, and the external forces $F_{\rm{ext}}^{i}$. We focus here on an external magnetic field as the only external force that acts on the AFM, giving
\begin{equation}
\label{eq:ExternalForces}
F_{\rm{ext}}^{i}=\frac{\rho\gamma}{a}\int dV[\rho\dot{\vec{H}}\cdot(\vec{n}\times\partial_{a_{i}}\vec{n})+L(\partial_{a_{i}}\vec{n}\cdot\partial^{i}\vec{H})]\,,
\end{equation}
where, in addition to the previously identified reactive force on the collective coordinates in AFMs due to time-varying magnetic fields,\cite{PhysRevLett.110.127208} we now identify a new force induced by a spatially inhomogeneous magnetic field. This force will necessarily influence how antiferromagnetic textures are excited by external magnetic fields.

\section{Domain wall dynamics}
\label{seq:DWMotion}
In this section, we return to systems where the order parameter varies along one dimension and discuss how the intrinsic magnetization influences the motion and detection of solitons in quasi-one-dimensional AFMs. Although the texture is assumed to vary only along one direction, the nearest neighbours to each spin may also exist along two (2D) or three (3D) axes. Later, we show how a N\'{e}el domain wall can be accelerated and controlled by a stationary and spatially inhomogeneous magnetic field.

\subsection{Antiferromagnetic domain walls}
In one-dimensional spin chains, the spatial variation of the staggered field $\vec{n}$ is constricted to the spin chain axis, $\hat{z}$. At equilibrium, the time evolution of the staggered field and the magnetization vanishes, $\dot{\vec{n}}=0$ and $\dot{\vec{m}}=0$, and Eq.~(\ref{eq:nlinresponse}) gives
\begin{equation}
 \vec{n}_{0}\times \left[A^{*}\partial_{z}^{2}\vec{n}_{0}+K_{z}(\vec{n}_{0}\cdot\hat{z})\hat{z}\right]\times\vec{n}_{0}=0\,.\label{eq:nequilibrium}
\end{equation}

\begin{figure}
\centering
\hspace{0.8cm}\includegraphics[scale=0.35]{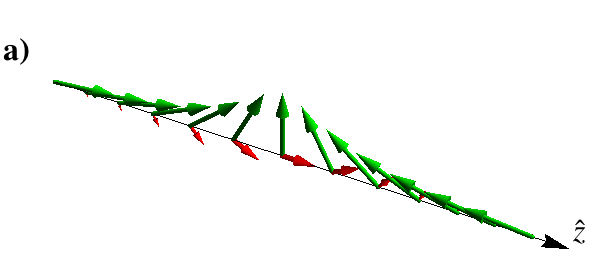}\\
\includegraphics[scale=0.35]{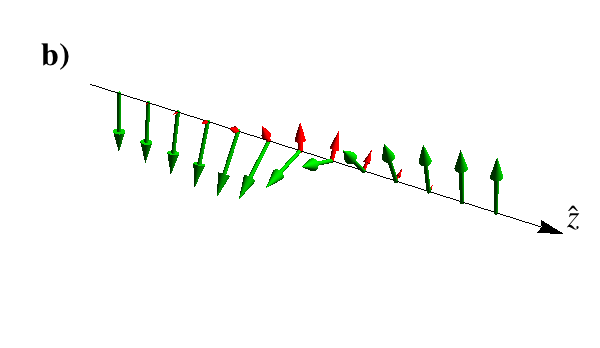}
\caption{Sketch of the intrinsic magnetization $\vec{m}(z)$ (red) (not to scale) of one-dimensional (\textbf{a}) N\'{e}el and (\textbf{b}) Bloch (not used in the calculations) domain walls in the order parameter $\vec{n}(z)$ (green). The equilibrium magnetization profile was calculated from Eq.~(\ref{eq:meq}). We note that the magnetization is so small that in a one-dimensional system, the domain wall spin must be treated quantum mechanically. However, for higher dimensional extended systems, the total spin of domain walls could be of appreciable size because the intrinsic magnetization is additive in the perpendicular directions.}
\label{fig:dwmagn}
\end{figure}

By introducing spherical coordinates for the normalized staggered vector field as  $\vec{n}_{0}(z)=\left[\sin\theta_{0}\cos\phi_{0},\sin\theta_{0}\sin\phi_{0},\cos\theta_{0}\right]$, a series of solutions for the above equation can be found from 
\begin{subequations}
\label{eqs:walkereqs}
\begin{eqnarray}
\partial_{z}\phi_{0}=0\,,\label{eq:phiequation}\\
\partial_{z}^{2}\theta_{0}=\frac{1}{\lambda^{2}}\sin\theta_{0}\cos\theta_{0}\,,
\label{eq:thetaequation}
\end{eqnarray}
\end{subequations}
where $\lambda=\sqrt{A^{*}/K_{z}}$. The trivial solution to Eqs.~(\ref{eqs:walkereqs}) is $\theta_{0}(z,t)\to0$, which corresponds to a homogeneous AFM where all the spins are polarized along the positive/negative $z$ direction. The excited state is given by $\theta_{0}=2\arctan[\exp(z/\lambda)]$, the Walker domain wall.\cite{schryer:5406} In this N\'{e}el configuration, $\sin\theta_{0}=\pm\mathrm{sech}(z/\lambda)$ and $\cos\theta_{0}=\pm\mathrm{tanh}(z/\lambda)$, which ensures that $\vec{n}_{0}^{2}=1$. $\lambda$ is the half-width of the domain wall. Inserting the results from the Heisenberg model, we find that the domain wall half-width $\lambda=d\sqrt{J/K}$ is given by a competition between the exchange and anisotropy energy scales, as expected.

The intrinsic magnetization associated with the antiferromagnetic domain wall at equilibrium is given by Eq.~(\ref{eq:meq}) when $\dot{\vec{n}}=0$:
\begin{equation}
\vec{m}_{0}=-\frac{L}{a}\partial_{z}\vec{n}_{0}=\pm\frac{d}{2\lambda}\begin{bmatrix}
\mathrm{sech}(z/\lambda)\tanh(z/\lambda)\cos\phi_{0}\\
\mathrm{sech}(z/\lambda)\tanh(z/\lambda)\sin\phi_{0}\\
-\mathrm{sech}^{2}(z/\lambda)
\end{bmatrix}\,,
\label{eq:equilibriummagnetization}
\end{equation}
where the sign determines whether the N\'{e}el domain wall is head-to-head or tail-to-tail. The magnetization profile of a head-to-head N\'{e}el domain wall and the profile of an antiferromagnetic Bloch domain wall are presented in Fig.~\ref{fig:dwmagn}. The total magnetic moment in the $z$ direction contained in a head-to-head domain wall configuration is\cite{PhysRevB.51.15062,PhysRevLett.74.1859}
\begin{equation}
M_{z}^{\mathrm{dw}}=\frac{S}{d}\int dz(\vec{m}_{0}\cdot\hat{z})=S\,,
\label{eq:domainwallmagnetization}
\end{equation}
This result demonstrates that domain walls in the antiferromagnetic order induce a finite magnetization proportional to the spatial derivative of the staggered field and that the direction of the magnetization depends crucially on the boundary conditions of the AFM, e.g., in the case of the N\'{e}el wall whether it is head-to-head or tail-to-tail. This result is intuitively easy to appreciate: because both edge spins (at an $\alpha$ and $\beta$ site) point in the same direction, the 180$^{\circ}$ twist turns the homogeneous spinless AFM into a spin-$S$ object. The domain wall is a non-linear excitation of the homogeneous AFM and carries the spin $S$. The creation of a domain wall can therefore be interpreted as a twisting of the homogeneous spinless AFM into a configuration with a finite spin $S$ that is located around the domain wall center.

A consequence of the intrinsic magnetization of domain walls in one-dimensional spin chains is that for AFMs with half-integer $S$, the ground state, which is doubly degenerate, occurs for stationary domain walls,\cite{PhysRevLett.74.1859,PhysRevB.56.8886} and not for precessing domain walls, as predicted in Ref.~\onlinecite{PhysRevLett.50.1153}. Another consequence is that the motion of domain walls in AFMs is equivalent to spin angular momentum transfer, as confirmed by Eq.~(\ref{eq:SpinCurrent}). The identification of antiferromagnetic domain walls as single-spin carriers may become important for future applications in antiferromagnetic spintronics.

\subsection{Domain wall motion}
We consider a (slowly) moving tail-to-tail domain wall profile $\vec{n}[z,a_{n}(t)]$ corresponding to the dynamic soliton solution $\theta(z,t)\to 2\mathrm{arctan}\{\mathrm{exp}[(z-r_{w}(t))/\lambda]\}$ and $\phi(t)\to \phi_{w}(t)$. The domain wall shape is assumed to be rigid, so that the temporal dynamics are held by the collective coordinates $\{a_{n}(t)\}\rightarrow\{\phi_{w}(t),r_{w}(t)\}$, the domain wall tilt angle with respect to the $x$-$z$ plane and the position of the domain wall center, respectively. Dissipation in AFMs is typically added in a phenomenological manner\cite{PhysRevLett.106.107206,PhysRevB.90.104406} and is naturally incorporated in the collective coordinate approach.\cite{PhysRevLett.110.127208} We add to the system a spatially varying magnetic field in the $\hat{z}$ direction, $\vec{H}=\{0,0,H_{z}(z)\}$. To the lowest order in the small external field and the velocities $\dot{\phi}_{w}$ and $\dot{r}_{w}$, we find that $\ddot{\phi}_{w}$ vanishes (although a constant precession $\dot{\phi}_{w}\neq 0$ is allowed in one-dimensional easy-axis systems) and that the domain wall center coordinate is accelerated according to
\begin{equation}
\label{eq:velocityHgradient}
\ddot{r}_{w}+\frac{a\alpha}{\rho}\dot{r}_{w}=-\frac{\gamma L}{\pi\rho\lambda}H_{z}^{\rm{int}}\,,
\end{equation} 
where $\alpha$ is the dimensionless Gilbert damping parameter of the AFM. Depending on the spatial profile of the magnetic field in the vicinity of the domain wall, the center coordinate will feel a force. The integrated magnetic field contribution is
\begin{equation}
\label{eq:HgradientIntegral}
H_{z}^{\rm{int}}=\int dz\left[\mathrm{sech}(\frac{z-r_{w}}{\lambda})\mathrm{tanh}(\frac{z-r_{w}}{\lambda})H_{z}(z)\right]\,,
\end{equation}
where any non-even profile $H_{z}(z)$ around the domain wall center coordinate $r_{w}$ gives rise to a finite acceleration of the domain wall. A homogeneous magnetic field does not accelerate the domain wall. In the steady state, the domain wall velocity saturates at $\dot{r}_{w}=\gamma LH_{z}^{\rm{int}}/(\pi a\alpha\lambda)$. We note that the domain wall velocity depends on the spatial distribution of the external magnetic field. This dependence opens up the possibility that nanoscale magnetic probes can accurately control the position of domain walls in, e.g., antiferromagnetic nanowires. In particular, a spatially constricted magnetic field can act as a potential well for the domain wall. In two-dimensional antiferromagnetic thin films, a spatially concentrated magnetic probe may attract spins from the edges of the AFM to form vortex states, see Sec.~\ref{sec:Discussion}.

\section{Numerical results}
\label{sec:Numerics}
To conceptually test the effect of a spatially inhomogeneous magnetic field on the dynamics of an antiferromagnetic domain wall, we have conducted numerical simulations of generalized versions of Eqs.~(\ref{eq:LLGn}) and (\ref{eq:LLGm}) in which we have phenomenologically included dissipation as in Ref.~\onlinecite{PhysRevLett.106.107206}. We write the equations of motion in dimensionless form by scaling the time axis by $\tilde{t}=\rho/K_{z}$ and the spatial axis by the nearest-neighbor distance $d$. We solve the dimensionless equations of motion using the numerical method of lines with an adaptive time control. The system size is $z\ \epsilon \left[-500,500\right]$ with the boundary conditions that $n_{z}(-500)=-1$ and $n_{z}(500)=1$.

Although domain walls in insulating AFMs, such as NiO, are approximately 100 nm wide,\cite{PhysRevLett.91.237205} we consider here the much shorter and more technologically important domain walls observed in antiferromagnetic Fe-monolayers on W(001)\cite{Bode:2006uq}, for which the geometric anisotropy is considerably larger. In such systems, the domain wall widths are only a few lattice spacings and the intrinsic magnetization is therefore relatively more important. For spin-1/2 particles, for which the anisotropy energy per atom is $2.4\ \mathrm{meV}$,\cite{PhysRevLett.94.087204} the time unit $\tilde{t}\approx 1$ ps, the velocity unit $\tilde{v}=d/\tilde{t}\approx 300\ \mathrm{ms^{-1}}$ and the external field unit $\tilde{h}=\rho\gamma/K_{z}\approx 0.3\ \mathrm{T}$.

\begin{figure}
\centering
\includegraphics[scale=0.48]{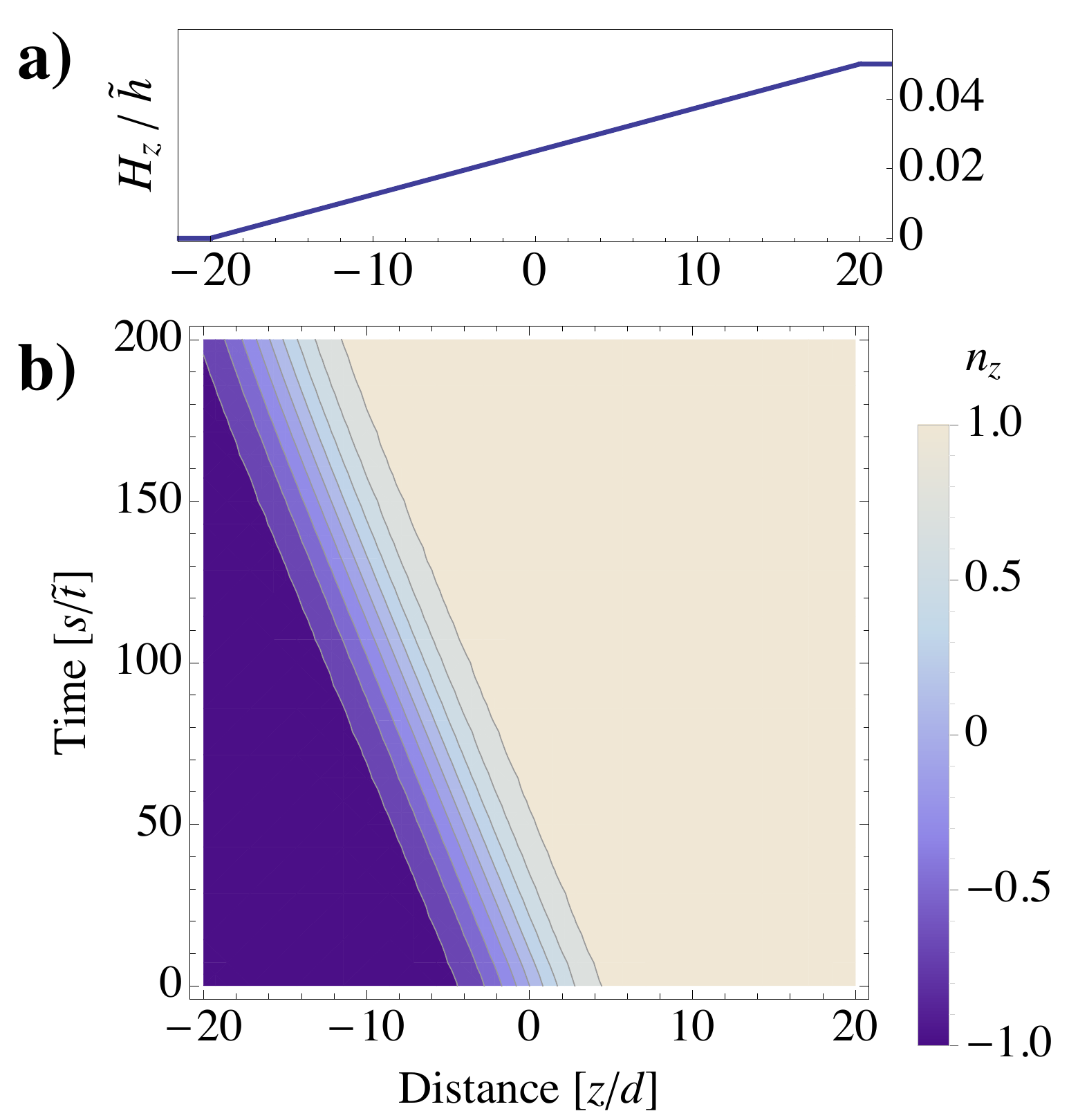}
\caption{(\textbf{a}) Magnetic field strength as a function of distance along the spin chain. The magnetic field has a constant gradient of approximately 0.4 mT per lattice constant. (\textbf{b}) Time evolution of an antiferromagnetic domain wall moved by the magnetic field gradient. For clarity only the region $z\ \epsilon [-25,25]$ is shown. The domain wall slows down due to the finite dissipation when it reaches the region of the homogeneous external field. The maximum magnetic field strength is $H^{\mathrm{max}}\approx 10\ \mathrm{mT}$, and the Gilbert damping constant is set to $\alpha=0.01$.}
\label{fig:dwcontourConst}
\end{figure}

\begin{figure}
\centering
\includegraphics[scale=0.48]{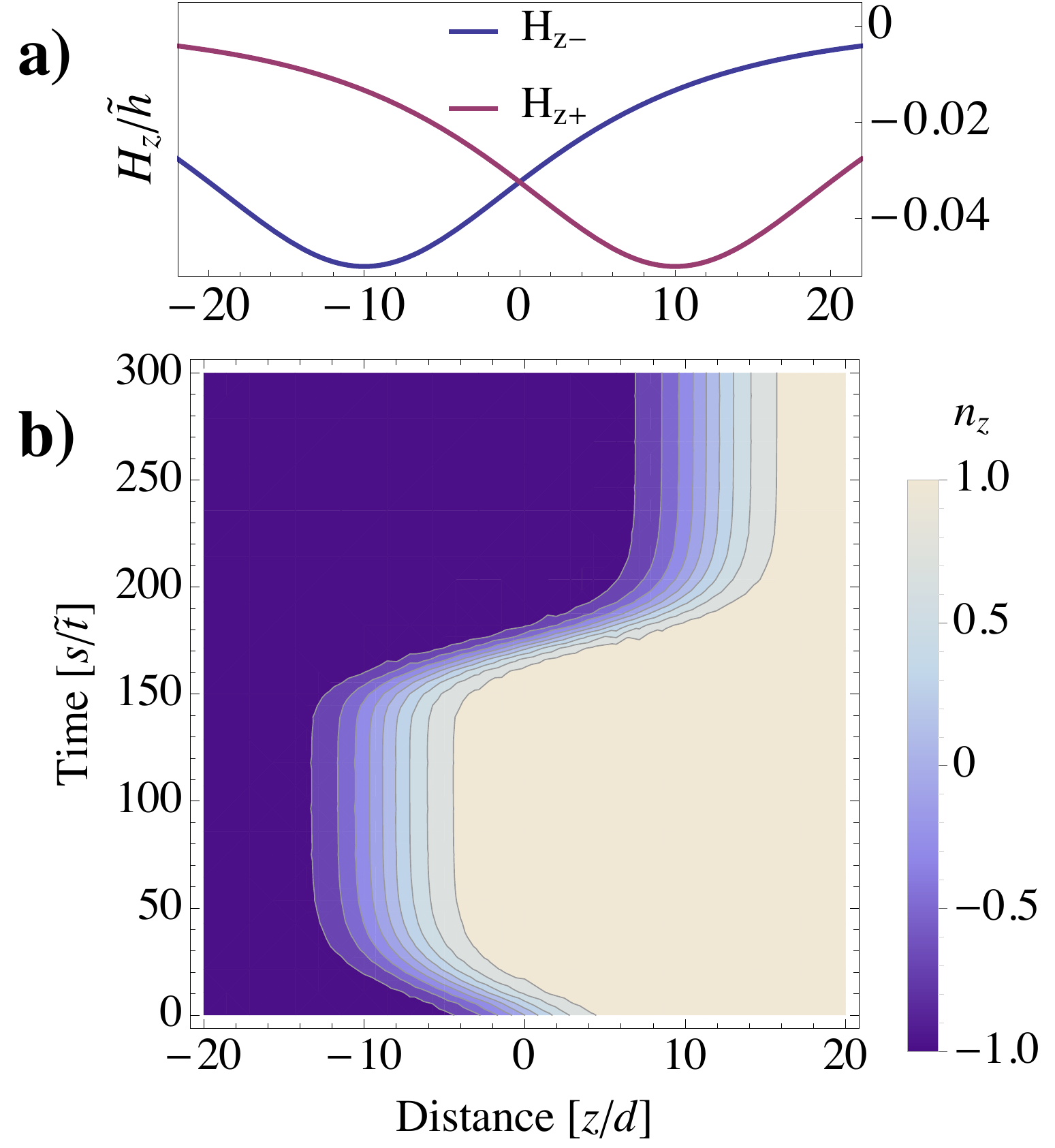}
\caption{(\textbf{a}) Magnetic field potential wells as a function of distance along the spin chain. (\textbf{b}) Time evolution of an antiferromagnetic domain wall controlled by the magnetic field potential wells. At $t=0$, the domain wall is attracted toward a potential well at $z_{-}=-10$ due to a spatially concentrated magnetic field in the $\hat{z}$ direction with the spatial profile $H_{z}=H_{0} \mathrm{sech}[(z-z_{-})/10]\hat{z}$. In the time interval $t=140\to 160$, the potential well to the left is turned off, and a similar magnetic field-induced potential well is turned on to the right at $z_{+}=10$.}
\label{fig:dwcontour}
\end{figure}

Fig.~\ref{fig:dwcontourConst} presents the motion of a domain wall with half-width $\lambda=4d$ due to a constant magnetic field gradient. Because the domain wall spin in this particular N\'{e}el domain wall is $-S$, the wall drifts toward lower magnetic fields to minimize its energy. The domain wall quickly reaches a steady-state velocity of approximately 50 ms$^{-1}$. Fig.~\ref{fig:dwcontour} presents how spatially concentrated magnetic fields can control and pin the position of the domain wall. By switching the pinning potential from the left to the right side of the domain wall, the position of the wall can be accurately controlled. The velocity of the center coordinate reaches more than 100 ms$^{-1}$, and the transition from the left to the right pinning potential occurs in less than 100 ps.

\section{Higher dimensional extended systems}
\label{sec:Discussion}
In this section, we discuss the possible experimental consequences for higher-dimensional textured systems, which typically extend in one or two perpendicular directions to the texture gradient axis. In such systems, the intrinsic magnetization can add up to a macroscopic number that is much larger than the spin on one atomic site. We also discuss the intrinsic magnetization of vortex states of the staggered order, which are two-dimensional analogs of the domain wall in the one-dimensional spin chain. At the end we briefly discuss the effects of pinning sites on the domain wall dynamics.

\subsection{Antiferromagnetic vortex states}
\begin{figure}
\centering
\includegraphics[scale=0.4]{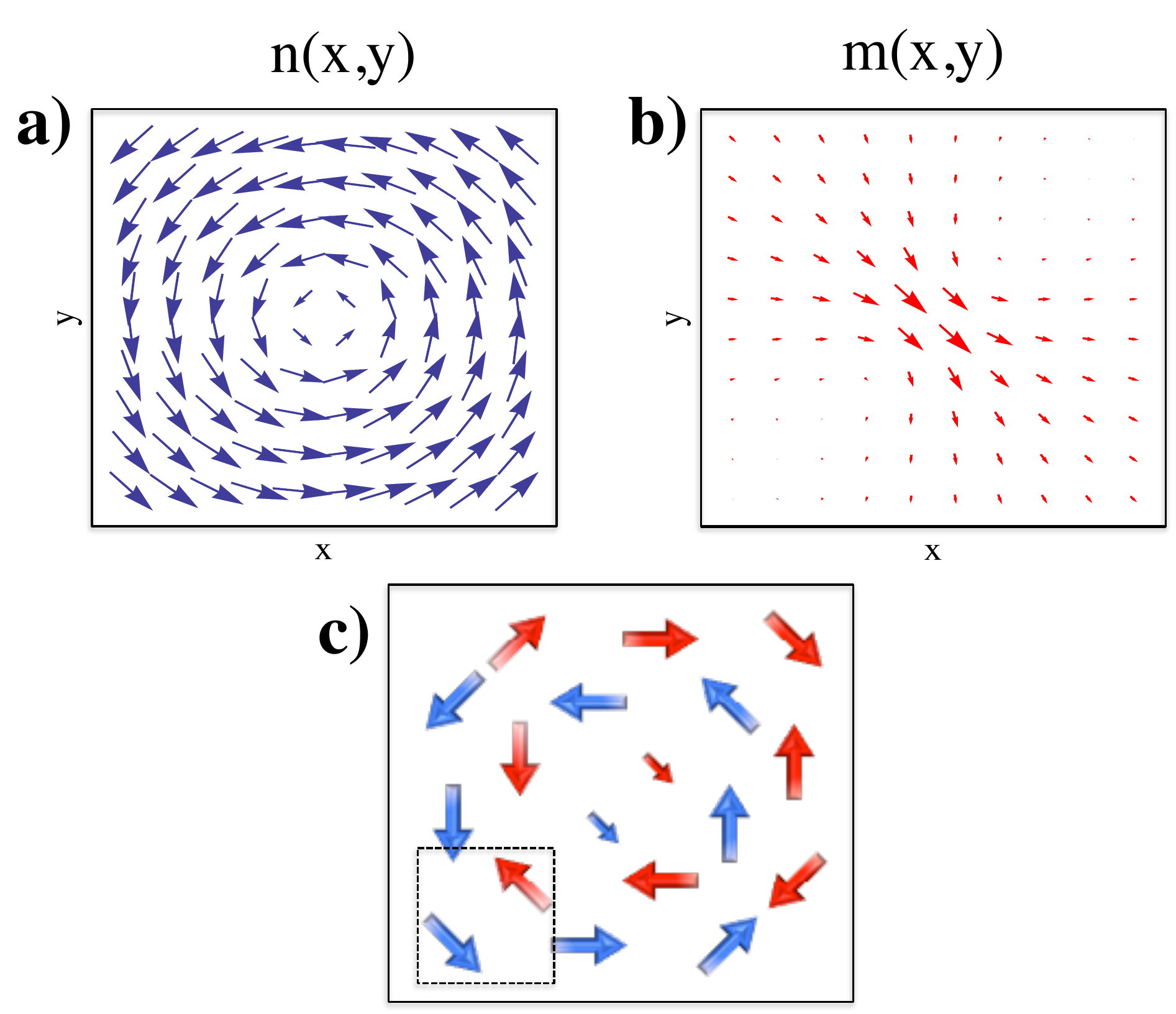}
\caption{\textbf{a}) Sketch of a 2-dimensional antiferromagnetic vortex structure in the staggered order parameter $\vec{n}(x,y)$ (blue vector field). The portions of the staggered field pointing in the perpendicular direction close to the vortex core have been omitted for clarity. \textbf{b}) The resulting magnetization density profile $\vec{m}(x,y)$ (red vector field, not to scale) of the vortex state, calculated from Eq.~(\ref{eq:meq}). This intrinsic magnetization can be interpreted as a rearrangement of the spins on the corners so that the center of the vortex structure acquires a finite spin. \textbf{c)} A simplified sketch of a vortex structure with six spins along each edge ordered in centered squared unit cells with $\alpha$ (blue arrows) and $\beta$ (red arrows) sites. Although the vortex structure in the continuous staggered field appears invariant under the rotation of an arbitrary angle around the vortex core, the underlying spin structure is only invariant under axis inversion, $[\hat{x},\hat{y}]\to[-\hat{x},-\hat{y}]$. The total integrated spin of the vortex structure is $S$, such as for the one-dimensional domain wall. The direction of this intrinsic spin is determined by the boundary conditions of the AFM.}
\label{fig:vortexmagn}
\end{figure}

For $D=2$ and in quasi-two-dimensional systems, such as antiferromagnetic thin films, non-trivial topological objects such as vortices\cite{Wu:2011yu} can form due to DM fields or external pinning. Fig.~\ref{fig:vortexmagn} shows the intrinsic magnetization $\vec{m}(x,y)$ associated with the spatially inhomogeneous staggered vector field of such a two-dimensional object. The magnetization profile is calculated from Eq.~(\ref{eq:meq}). We note that the intrinsic spin of the vortex structure can be interpreted as a twisting of the spins in the homogeneous spinless AFM  induced by spin rotations on the corners into a state with a finite spin located around the vortex core. The staggered vector field $\vec{n}(x,y)$ of this type of vortex structure is rotationally invariant around the vortex core along an axis normal to the $x$-$y$ plane. The underlying spin structure, however, is not rotationally invariant, which is captured by the finite magnetization density $\vec{m}(x,y)$ of the vortex. The total spin of the vortex is $S$, as in the case of a domain wall, and the direction of the intrinsic spin depends crucially on the boundary conditions of the AFM, e.g., induced via exchange bias pinning to ferromagnetic neighbors.

The topological term in the effective Lagrangian density (\ref{eq:efflagrange}) for the staggered vector field $\vec{n}$ can possibly indirectly influence the dynamics of two-dimensional objects in the order parameter such as vortices or skyrmions. However, the complex two-dimensional dynamics of such topological objects are beyond the scope of this article and will not be discussed further.

\subsection{Extended domain walls in 2D and 3D}

\begin{figure}
\centering
\includegraphics[scale=0.40]{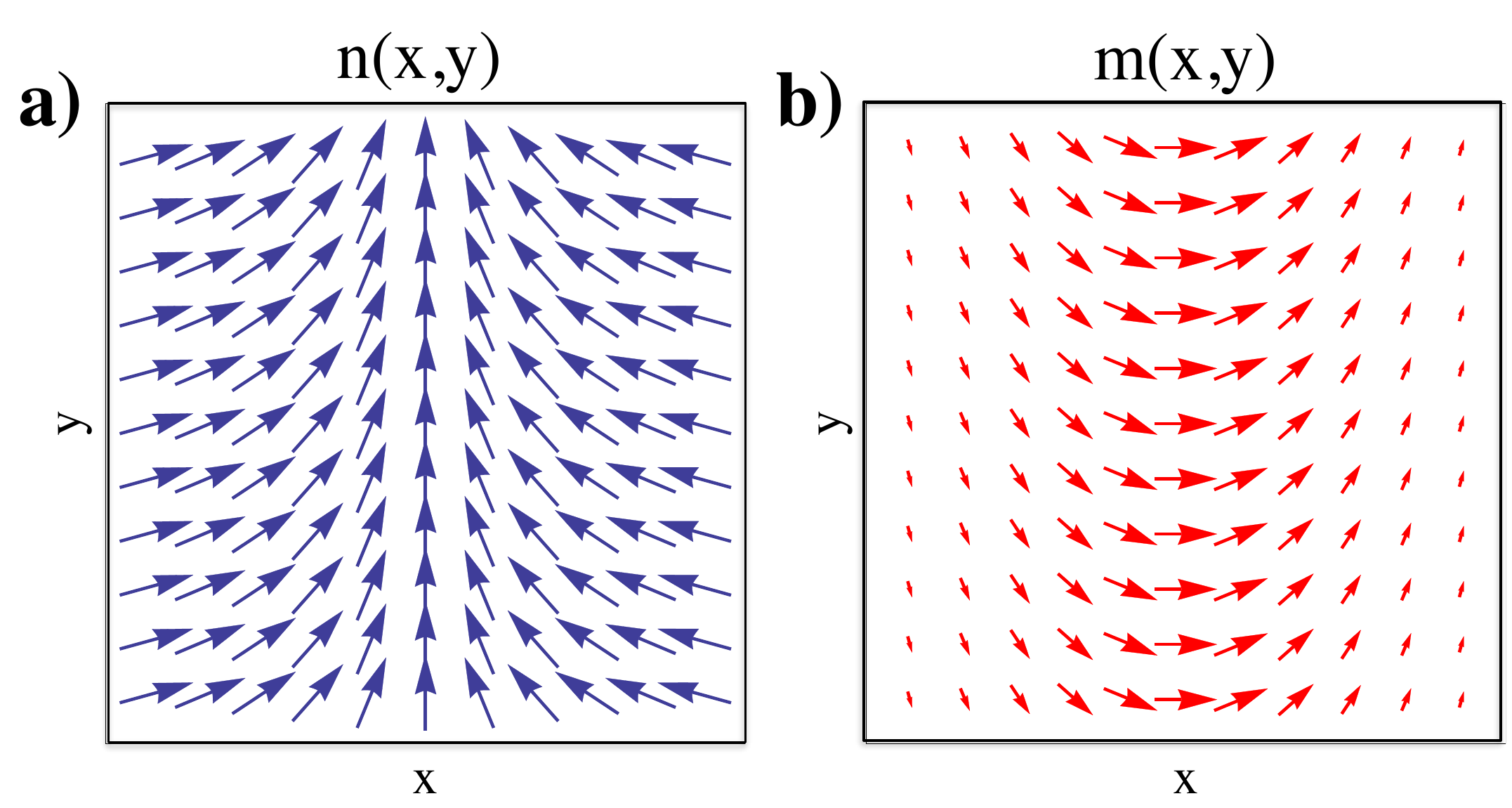}
\caption{\textbf{a}) Sketch of a domain wall in the staggered order parameter $\vec{n}(x,y)$ (blue vector field) in a two-dimensional AFM, e.g., a nanostrip. The domain wall configuration is repeated in the perpendicular direction. \textbf{b}) The magnetization vector field $\vec{m}(z,y)$ (red vector field, not to scale) calculated from Eq.~(\ref{eq:meq}). Each spin chain along the horizontal direction will contribute the spin $S\hat{x}$ to the total spin of the two-dimensional domain wall structure.}
\label{fig:dw2dmagn}
\end{figure}

Because the intrinsic spin of one-dimensional domain walls and two-dimensional vortices totals no more than the spin on a single atomic lattice site $S$, it is unlikely that the intrinsic magnetization associated with these antiferromagnetic textures can be reliably detected in the near future. Furthermore, to predict the correct excitation scheme of antiferromagnetic solitons, the intrinsic spin must be treated quantum mechanically because quantum fluctuations become important.\cite{PhysRevLett.74.1859} In higher-dimensional systems such as thin films or bulk AFMs, however, domain walls are not purely one-dimensional objects. Although the order parameter can be defined as varying along one axis only, the nearest neighbors of each spin can exist along two (2D) or three (3D) perpendicular axes. In such systems, the intrinsic magnetization of the domain wall accumulates over the total number of spin chains that constitute the domain wall structure. An example of the intrinsic magnetization of such an extended domain wall structure in, e.g., a nanostrip is presented in Fig.~\ref{fig:dw2dmagn}.

In bulk AFMs with domain structures in the order parameter, the intrinsic magnetization forms planes along the domain boundaries. The total spin of these magnetization planes can be of appreciable size. In addition, for synthetic antiferromagnetic superlattices, in which the magnetization of each single ferromagnetic layer is much larger than $S$, the intrinsic magnetization associated with magnetic textures is accordingly larger and may be detectable.\cite{PapanicolaouJPCM1998}

\subsection{Effect of pinning sites on domain wall dynamics}
Pinning sites for domain walls can arise from impurities or crystal defects in the underlying lattice of AFMs. Although several studies have found that pinning effects in AFMs are small,\cite{PhysRevB.44.7413,Shpyrko:2007fk,logan:192405} dislocations and impurities diffuse the effects on a single spin. In quasi-one-dimensional spin chains the introduction of a single impurity atom can be enough to destroy long-ranged antiferromagnetic order and domain wall configurations. From such a perspective, the scenario studied in Sec.~\ref{sec:Numerics} requires a perfect spin chain in strictly one-dimensional systems. However, because three-dimensional domain boundaries are typically sums of many one-dimensional spin chains, we expect the effects of pinning from impurities to be significantly smaller for domain wall systems that extend in the perpendicular directions than for one-dimensional spin chains.

\section{Conclusion}
\label{sec:Conclusion}
Starting from the discrete Heisenberg Hamiltonian with antiferromagnetic exchange coupling and easy-axis anisotropy, we have rederived the continuum limit of the free energy functional in the exchange approximation and conclusively shown that textures in the antiferromagnetic order exhibit intrinsic magnetization. In recent effective models for the dynamics of the antiferromagnetic order parameter, this intrinsic magnetization has been mostly disregarded. By comparing the Hamiltonian approach that we apply in this article with a commonly used alternative parametrization procedure called Haldane's mapping, we have shown that the continuum fields of the two parametrization procedures have crucially different physical interpretations. As a result, the intrinsic magnetization can be easily missed in continuum models based on Haldane's mapping.

We have demonstrated that parity-breaking terms in the energy functional influence the dynamics of textured AFMs affected by external forces that couple directly to the intrinsic magnetization. For extended domain walls in 2D/3D the influence of the intrinsic magnetization on the texture dynamics goes beyond that of the quantum effects observed for one-dimensional spin chains. By utilizing the method of collective coordinates, we have shown that a spatially inhomogeneous magnetic field represents a reactive force on antiferromagnetic textures and can move a domain wall in an antiferromagnetic nanowire. This effect is directly linked to the intrinsic magnetization of the domain wall. Numerical simulations of the coupled equations of motion for the staggered field and the magnetization field confirmed that a spatially inhomogeneous magnetic field can act as a potential well for the domain wall. Finally, we have discussed how the intrinsic magnetization of antiferromagnetic textures, which for one-dimensional domain walls is not larger than the spin on one atomic site, can be experimentally exploited in 2D and 3D. In such higher-dimensional real systems the intrinsic magnetization accumulates in the perpendicular directions and the total spin can, therefore, be of appreciable size and may be detectable.

\begin{acknowledgments}
We thank Hans Skarsv\aa g, Rembert A.~Duine, and John-Ove Fj\ae restad for valuable discussions.
J.~L.~acknowledges support from the Outstanding Academic Fellows programme at NTNU and the Norwegian Research Council Grant No.~205591 and Grant No.~216700.
\end{acknowledgments}

\appendix

\section{Energy functional for $D>1$}
\label{app:2d}

In this Appendix, we expand our calculation of the free energy functional of AFMs to 2 and 3 dimensions to disclose the form of the parity-breaking term in higher dimensions. For $D=2$, we use the centered rectangular unit cell, with two sublattices within each unit cell. Starting with Eq.~(\ref{eq:HeisenbergHamiltonian}), we now assume that $\boldsymbol{\alpha}$ and $\boldsymbol{\beta}$ are two-dimensional vectors and that the coordinate pair $\{i,j\}$ unambiguously defines all antiferromagnetic unit cells. Next, we define
\begin{subequations}
\label{eqs:2ddefinition}
\begin{eqnarray}
\vec{m}_{i,j}&=&\frac{\vec{S}_{\alpha}^{i,j}+\vec{S}_{\beta}^{i,j}}{2S}\,,
\label{eq:mparameter2d}\\
\vec{l}_{i,j}&=&\frac{\vec{S}_{\alpha}^{i,j}-\vec{S}_{\beta}^{i,j}}{2S}\,,
\label{eq:nparameter2d}\\
\vec{S}_{\alpha}^{i,j}&=&S(\vec{m}_{i,j}+\vec{l}_{i,j})\,,
\label{eq:salpha2d}\\
\vec{S}_{\beta}^{i,j}&=&S(\vec{m}_{i,j}-\vec{l}_{i,j})\,,
\label{eq:sbeta2d}
\end{eqnarray}
\end{subequations}
where we must take into account the equivalence of interchanging $\vec{l}_{i}\rightarrow-\vec{l}_{i}$, such as in the one-dimensional derivation. The Heisenberg Hamiltonian can be written as a sum over antiferromagnetic unit cells in the perpendicular $i$ and $j$ directions
\begin{eqnarray}
 H_{\rm{2D}} & =& J S^{2}\sum_{i,j}^{N-1,N-1}(\vec{m}_{i,j}-\vec{l}_{i,j})[(\vec{m}_{i,j}+\vec{l}_{i,j})\nonumber\\
& &+(\vec{m}_{i+1,j}+\vec{l}_{i+1,j})+(\vec{m}_{i,j+1}+\vec{l}_{i,j+1})
\nonumber\\
& &+(\vec{m}_{i+1,j+1}+\vec{l}_{i+1,j+1})]\nonumber\\
& &-K S^{2}\sum_{i,j}^{N,N} \left[(\vec{m}_{i,j}+\vec{l}_{i,j})_{z}^{2}+(\vec{m}_{i,j}-\vec{l}_{i,j})_{z}^{2}\right]\,,
\label{eq:Heisenberg2dRewritten}
\end{eqnarray}
where we have disregarded a small energy contribution from the edge spins like in Sec.~\ref{sec:Theory1}. Eq.~(\ref{eq:Heisenberg2dRewritten}) is a sum over the $N_{n}=4$ nearest-neighbor exchange couplings and the anisotropy energies for each antiferromagnetic unit cell. We use the identities $2\vec{m}_{i,j}\vec{m}_{i+1,j}=\vec{m}_{i,j}^2+\vec{m}_{i+1,j}^2-(\vec{m}_{i+1,j}-\vec{m}_{i,j})^2$ and $(\vec{l}_{i,j}\vec{m}_{i+1,j}-\vec{m}_{i,j}\vec{l}_{i+1,j})=\vec{l}_{i,j}(\vec{m}_{i+1,j}-\vec{m}_{i,j})-\vec{m}_{i,j}(\vec{l}_{i+1,j}-\vec{l}_{i,j})$ etc.~to rewrite Eq.~(\ref{eq:Heisenberg2dRewritten}) to
\begin{eqnarray}
H_{\rm{2D}} &= & N_{n}J S^{2}\sum_{i,j}^{N,N}(\vec{m}_{i,j}^2-\vec{l}_{i,j}^2)\nonumber\\
& &+\frac{J S^{2}}{2}\sum_{i,j}^{N-1,N-1}[(\vec{l}_{i+1,j}-\vec{l}_{i,j})^2+(\vec{l}_{i,j+1}-\vec{l}_{i,j})^2\nonumber\\
& &+(\vec{l}_{i+1,j+1}-\vec{l}_{i,j})^2-(\vec{m}_{i+1,j}-\vec{m}_{i,j})^2\nonumber\\
& & -(\vec{m}_{i,j+1}-\vec{m}_{i,j})^2-(\vec{m}_{i+1,j+1}-\vec{m}_{i,j})^2]\nonumber\\
& &+ JS^2\sum_{i,j}^{N-1,N-1}[\vec{m}_{i,j}(\vec{l}_{i+1,j}+\vec{l}_{i,j+1}+\vec{l}_{i+1,j+1}-3\vec{l}_{i,j})\nonumber\\
& &-\vec{l}_{i,j}(\vec{m}_{i+1,j}+\vec{m}_{i,j+1}+\vec{m}_{i+1,j+1}-3\vec{m}_{i,j})]\nonumber\\
& &-2K S^{2}\sum_{i,j}^{N,N}(m_{i,j,z}^2+l_{i,j,z}^{2})\,.
\label{eq:Heisenberg1dRewritten2}
\end{eqnarray} 
To make the transition to the continuum limit, we define the derivatives in the linear approximation 
\begin{subequations}
\label{eqs:nderivatives}
\begin{eqnarray}
\label{eq:ContinuumLimit}
\lim_{|\Delta_{i}|\rightarrow 0}\sum_{i,j}(\vec{l}_{i+1,j}-\vec{l}_{i,j}) &\approx &  \frac{1}{V}\int[\mathcal{J}(\vec{l})\boldsymbol{\Delta_{i}}]\rm{dV}\,,\\
\lim_{|\Delta_{j}|\rightarrow 0}\sum_{i,j}(\vec{l}_{i,j+1}-\vec{l}_{i,j}) & \approx & \frac{1}{V}\int[\mathcal{J}(\vec{l})\boldsymbol{\Delta_{j}}]\rm{dV}\,,\\
\lim_{|\Delta_{i(j)}|\rightarrow 0}\sum_{i,j}(\vec{l}_{i+1,j+1}-\vec{l}_{i,j}) & \approx &  
 \frac{1}{V}\int[\mathcal{J}(\vec{l})\boldsymbol{\Delta_{i}}+\mathcal{J}(\vec{l})\boldsymbol{\Delta_{j}}]\rm{dV}\,,\nonumber\\
&& 
\end{eqnarray}
\end{subequations}
where $\mathcal{J}(\vec{l})$ is the Jacobian matrix of the vector field $\vec{l}$, $\boldsymbol{\Delta_{i(j)}}$ is a vector between unit cells in the $\hat{i}(\hat{j})$ direction, and $V$ is the volume of the unit cell. For the centered squared unit cell, $|\boldsymbol{\Delta_{i}}|=|\boldsymbol{\Delta_{j}}|\equiv\Delta$ and $V=\Delta^{2}$. We define similar derivatives as in Eqs.~(\ref{eqs:nderivatives}) for the magnetization field $\vec{m}$.

The procedure is analogous when including a third dimension, e.g., for a body-centered cubic unit cell, repeating the above calculation with $\{i,j\}\rightarrow\{i,j,k\}$. Apart from a constant contribution, the resulting free energy density for AFMs in dimensions $D>1$, defined here as $H_{\mathrm{2(3)D}}=\int(dV/V)\mathcal{H}_{\mathrm{2(3)D}}$, is given by
\begin{eqnarray}
\mathcal{H}_{\rm{2(3)D}} & =&  J S^{2}N_{n}\{2\vec{m}^2+\frac{1}{2}\sum_{i}\Delta_{i}^{2}[(\partial_{i}\vec{l})^2-(\partial_{i}\vec{m})^2]\nonumber\\
& & +\frac{1}{4}\sum_{i\neq j}\Delta_{i}\Delta_{j}(\partial_i\vec{n}\cdot\partial_{j}\vec{n}-\partial_{i}\vec{m}\cdot\partial_{j}\vec{m})\nonumber\\
& & +\frac{1}{2}\sum_{i}\Delta_{i}(\vec{m}\cdot\partial_{i}\vec{l}-\vec{l}\cdot\partial_{i}\vec{m})\}\nonumber\\
& &-K S^{2}[(\vec{l}\cdot\hat{z})^{2}+(\vec{m}\cdot\hat{z})^{2}]\,,
\label{eq:EnergyDensity2D}
\end{eqnarray}
where we may define $i$ and $j$ to run over perpendicular directions $\{x,y,z\}$. The sum over first order derivatives arises from the relation $\mathcal{J}(\vec{l})\boldsymbol{\Delta}=\sum_{i}\Delta_{i}\partial_{j}(\vec{l})$, where $\boldsymbol{\Delta}=\{\Delta_{i},\Delta_{j},\Delta_k\}$.

By considering squared or cubic lattices, $\Delta=2d/\sqrt{D}$ and $d$ is the nearest-neighbor distance. We can express the free energy density in the exchange approximation, $|K|\ll|J|$, as

\begin{eqnarray}
\mathcal{H}_{\rm{2(3)D}} &=& \frac{a}{2}\vec{m}^2+\frac{A}{2}\left[\sum_{i}(\partial_{i}\vec{n})^2+\frac{1}{2}\sum_{i\neq j}\partial_{i}\vec{n}\cdot\partial_{j}\vec{n}\right]\nonumber\\
& & + L\sum_{i}(\vec{m}\cdot\partial_{i}\vec{n})-\frac{K_{z}}{2}(\vec{n}\cdot\hat{z})^{2}\,,
\label{eq:EnergyDensity2DExchApprox}
\end{eqnarray}
where $a=4N_{n} J S^{2}$, $A=N_{n}\Delta^{2} J S^{2}/2$, $L=N_{n}\Delta J S^{2}$, $K_{z}=2 K S^{2}$, and $N_{n}$ is the number of nearest neighbors. 

In antiferromagnetic materials in which the exchange energy is anisotropic due to, e.g., more complicated unit cells, Eq.~(\ref{eq:EnergyDensity2DExchApprox}) can still be used, although in this case $a$, $A$, and $L$ must be treated as tensors.


\begin{thebibliography}{59}%
\makeatletter
\providecommand \@ifxundefined [1]{%
 \@ifx{#1\undefined}
}%
\providecommand \@ifnum [1]{%
 \ifnum #1\expandafter \@firstoftwo
 \else \expandafter \@secondoftwo
 \fi
}%
\providecommand \@ifx [1]{%
 \ifx #1\expandafter \@firstoftwo
 \else \expandafter \@secondoftwo
 \fi
}%
\providecommand \natexlab [1]{#1}%
\providecommand \enquote  [1]{``#1''}%
\providecommand \bibnamefont  [1]{#1}%
\providecommand \bibfnamefont [1]{#1}%
\providecommand \citenamefont [1]{#1}%
\providecommand \href@noop [0]{\@secondoftwo}%
\providecommand \href [0]{\begingroup \@sanitize@url \@href}%
\providecommand \@href[1]{\@@startlink{#1}\@@href}%
\providecommand \@@href[1]{\endgroup#1\@@endlink}%
\providecommand \@sanitize@url [0]{\catcode `\\12\catcode `\$12\catcode
  `\&12\catcode `\#12\catcode `\^12\catcode `\_12\catcode `\%12\relax}%
\providecommand \@@startlink[1]{}%
\providecommand \@@endlink[0]{}%
\providecommand \url  [0]{\begingroup\@sanitize@url \@url }%
\providecommand \@url [1]{\endgroup\@href {#1}{\urlprefix }}%
\providecommand \urlprefix  [0]{URL }%
\providecommand \Eprint [0]{\href }%
\@ifxundefined \urlstyle {%
  \providecommand \doi  [0]{\begingroup \@sanitize@url \@doi}%
  \providecommand \@doi [1]{\endgroup \@@startlink {\doibase
  #1}doi:\discretionary {}{}{}#1\@@endlink }%
}{%
  \providecommand \doi  [0]{doi:\discretionary{}{}{}\begingroup
  \urlstyle{rm}\Url }%
}%
\providecommand \doibase [0]{http://dx.doi.org/}%
\providecommand \Doi [0]{\begingroup \@sanitize@url \@Doi }%
\providecommand \@Doi  [1]{\endgroup\@@startlink{\doibase#1}\@@Doi}%
\providecommand \@@Doi [1]{#1\@@endlink}%
\providecommand \selectlanguage [0]{\@gobble}%
\providecommand \bibinfo  [0]{\@secondoftwo}%
\providecommand \bibfield  [0]{\@secondoftwo}%
\providecommand \translation [1]{[#1]}%
\providecommand \BibitemOpen [0]{}%
\providecommand \bibitemStop [0]{}%
\providecommand \bibitemNoStop [0]{.\EOS\space}%
\providecommand \EOS [0]{\spacefactor3000\relax}%
\providecommand \BibitemShut  [1]{\csname bibitem#1\endcsname}%
\bibitem [{\citenamefont {MacDonald}\ and\ \citenamefont
  {Tsoi}(2011)}]{MacDonald13082011}%
  \BibitemOpen
  \bibfield  {author} {\bibinfo {author} {\bibfnamefont {A.~H.}\ \bibnamefont
  {MacDonald}}\ and\ \bibinfo {author} {\bibfnamefont {M.}~\bibnamefont
  {Tsoi}},\ }\href
  {http://rsta.royalsocietypublishing.org/content/369/1948/3098.abstract}
  {\bibfield  {journal} {\bibinfo  {journal} {Phil. Trans. R. Soc. A},\
  }\textbf {\bibinfo {volume} {369}},\ \bibinfo {pages} {3098} (\bibinfo {year}
  {2011})}\BibitemShut {NoStop}%
\bibitem [{\citenamefont {Gomonay}\ and\ \citenamefont
  {Loktev}(2014)}]{GomonayLTP2013}%
  \BibitemOpen
  \bibfield  {author} {\bibinfo {author} {\bibfnamefont {E.~V.}\ \bibnamefont
  {Gomonay}}\ and\ \bibinfo {author} {\bibfnamefont {V.~M.}\ \bibnamefont
  {Loktev}},\ }\Doi {http://dx.doi.org/10.1063/1.4862467} {\bibfield  {journal}
  {\bibinfo  {journal} {Low Temperature Physics},\ }\textbf {\bibinfo {volume}
  {40}},\ \bibinfo {pages} {17} (\bibinfo {year} {2014})}\BibitemShut {NoStop}%
\bibitem [{\citenamefont {Park}\ \emph {et~al.}(2011)\citenamefont {Park},
  \citenamefont {Wunderlich}, \citenamefont {Mart{\'\i}}, \citenamefont
  {Hol{\'y}}, \citenamefont {Kurosaki}, \citenamefont {Yamada}, \citenamefont
  {Yamamoto}, \citenamefont {Nishide}, \citenamefont {Hayakawa}, \citenamefont
  {Takahashi}, \citenamefont {Shick},\ and\ \citenamefont
  {Jungwirth}}]{Park:2011rt}%
  \BibitemOpen
  \bibfield  {author} {\bibinfo {author} {\bibfnamefont {B.~G.}\ \bibnamefont
  {Park}}, \bibinfo {author} {\bibfnamefont {J.}~\bibnamefont {Wunderlich}},
  \bibinfo {author} {\bibfnamefont {X.}~\bibnamefont {Mart{\'\i}}}, \bibinfo
  {author} {\bibfnamefont {V.}~\bibnamefont {Hol{\'y}}}, \bibinfo {author}
  {\bibfnamefont {Y.}~\bibnamefont {Kurosaki}}, \bibinfo {author}
  {\bibfnamefont {M.}~\bibnamefont {Yamada}}, \bibinfo {author} {\bibfnamefont
  {H.}~\bibnamefont {Yamamoto}}, \bibinfo {author} {\bibfnamefont
  {A.}~\bibnamefont {Nishide}}, \bibinfo {author} {\bibfnamefont
  {J.}~\bibnamefont {Hayakawa}}, \bibinfo {author} {\bibfnamefont
  {H.}~\bibnamefont {Takahashi}}, \bibinfo {author} {\bibfnamefont {A.~B.}\
  \bibnamefont {Shick}}, \ and\ \bibinfo {author} {\bibfnamefont
  {T.}~\bibnamefont {Jungwirth}},\ }\href {http://dx.doi.org/10.1038/nmat2983}
  {\bibfield  {journal} {\bibinfo  {journal} {Nat. Mater.},\ }\textbf {\bibinfo
  {volume} {10}},\ \bibinfo {pages} {347} (\bibinfo {year} {2011})}\BibitemShut
  {NoStop}%
\bibitem [{\citenamefont {Mart\'\i}\ \emph {et~al.}(2012)\citenamefont
  {Mart\'\i}, \citenamefont {Park}, \citenamefont {Wunderlich}, \citenamefont
  {Reichlov\'a}, \citenamefont {Kurosaki}, \citenamefont {Yamada},
  \citenamefont {Yamamoto}, \citenamefont {Nishide}, \citenamefont {Hayakawa},
  \citenamefont {Takahashi},\ and\ \citenamefont
  {Jungwirth}}]{PhysRevLett.108.017201}%
  \BibitemOpen
  \bibfield  {author} {\bibinfo {author} {\bibfnamefont {X.}~\bibnamefont
  {Mart\'\i}}, \bibinfo {author} {\bibfnamefont {B.~G.}\ \bibnamefont {Park}},
  \bibinfo {author} {\bibfnamefont {J.}~\bibnamefont {Wunderlich}}, \bibinfo
  {author} {\bibfnamefont {H.}~\bibnamefont {Reichlov\'a}}, \bibinfo {author}
  {\bibfnamefont {Y.}~\bibnamefont {Kurosaki}}, \bibinfo {author}
  {\bibfnamefont {M.}~\bibnamefont {Yamada}}, \bibinfo {author} {\bibfnamefont
  {H.}~\bibnamefont {Yamamoto}}, \bibinfo {author} {\bibfnamefont
  {A.}~\bibnamefont {Nishide}}, \bibinfo {author} {\bibfnamefont
  {J.}~\bibnamefont {Hayakawa}}, \bibinfo {author} {\bibfnamefont
  {H.}~\bibnamefont {Takahashi}}, \ and\ \bibinfo {author} {\bibfnamefont
  {T.}~\bibnamefont {Jungwirth}},\ }\Doi {10.1103/PhysRevLett.108.017201}
  {\bibfield  {journal} {\bibinfo  {journal} {Phys. Rev. Lett.},\ }\textbf
  {\bibinfo {volume} {108}},\ \bibinfo {pages} {017201} (\bibinfo {year}
  {2012})}\BibitemShut {NoStop}%
\bibitem [{\citenamefont {Marti}\ \emph {et~al.}(2014)\citenamefont {Marti},
  \citenamefont {Fina}, \citenamefont {Frontera}, \citenamefont {Liu},
  \citenamefont {Wadley}, \citenamefont {He}, \citenamefont {Paull},
  \citenamefont {Clarkson}, \citenamefont {Kudrnovsk\'{y}}, \citenamefont
  {Turek}, \citenamefont {Kune\v{s}}, \citenamefont {Yi}, \citenamefont {Chu},
  \citenamefont {Nelson}, \citenamefont {You}, \citenamefont {Arenholz},
  \citenamefont {Salahuddin}, \citenamefont {Fontcuberta}, \citenamefont
  {Jungwirth},\ and\ \citenamefont {Ramesh}}]{Marti:2014xy}%
  \BibitemOpen
  \bibfield  {author} {\bibinfo {author} {\bibfnamefont {X.}~\bibnamefont
  {Marti}}, \bibinfo {author} {\bibfnamefont {I.}~\bibnamefont {Fina}},
  \bibinfo {author} {\bibfnamefont {C.}~\bibnamefont {Frontera}}, \bibinfo
  {author} {\bibfnamefont {J.}~\bibnamefont {Liu}}, \bibinfo {author}
  {\bibfnamefont {P.}~\bibnamefont {Wadley}}, \bibinfo {author} {\bibfnamefont
  {Q.}~\bibnamefont {He}}, \bibinfo {author} {\bibfnamefont {R.~J.}\
  \bibnamefont {Paull}}, \bibinfo {author} {\bibfnamefont {J.~D.}\ \bibnamefont
  {Clarkson}}, \bibinfo {author} {\bibfnamefont {J.}~\bibnamefont
  {Kudrnovsk\'{y}}}, \bibinfo {author} {\bibfnamefont {I.}~\bibnamefont
  {Turek}}, \bibinfo {author} {\bibfnamefont {J.}~\bibnamefont {Kune\v{s}}},
  \bibinfo {author} {\bibfnamefont {D.}~\bibnamefont {Yi}}, \bibinfo {author}
  {\bibfnamefont {J.-H.}\ \bibnamefont {Chu}}, \bibinfo {author} {\bibfnamefont
  {C.~T.}\ \bibnamefont {Nelson}}, \bibinfo {author} {\bibfnamefont
  {L.}~\bibnamefont {You}}, \bibinfo {author} {\bibfnamefont {E.}~\bibnamefont
  {Arenholz}}, \bibinfo {author} {\bibfnamefont {S.}~\bibnamefont
  {Salahuddin}}, \bibinfo {author} {\bibfnamefont {J.}~\bibnamefont
  {Fontcuberta}}, \bibinfo {author} {\bibfnamefont {T.}~\bibnamefont
  {Jungwirth}}, \ and\ \bibinfo {author} {\bibfnamefont {R.}~\bibnamefont
  {Ramesh}},\ }\href {http://dx.doi.org/10.1038/nmat3861} {\bibfield  {journal}
  {\bibinfo  {journal} {Nat. Mater.},\ }\textbf {\bibinfo {volume} {13}},\
  \bibinfo {pages} {367} (\bibinfo {year} {2014})}\BibitemShut {NoStop}%
\bibitem [{\citenamefont {Wang}\ \emph {et~al.}(2014)\citenamefont {Wang},
  \citenamefont {Seinige}, \citenamefont {Cao}, \citenamefont {Zhou},
  \citenamefont {Goodenough},\ and\ \citenamefont {Tsoi}}]{PhysRevX.4.041034}%
  \BibitemOpen
  \bibfield  {author} {\bibinfo {author} {\bibfnamefont {C.}~\bibnamefont
  {Wang}}, \bibinfo {author} {\bibfnamefont {H.}~\bibnamefont {Seinige}},
  \bibinfo {author} {\bibfnamefont {G.}~\bibnamefont {Cao}}, \bibinfo {author}
  {\bibfnamefont {J.~S.}\ \bibnamefont {Zhou}}, \bibinfo {author}
  {\bibfnamefont {J.~B.}\ \bibnamefont {Goodenough}}, \ and\ \bibinfo {author}
  {\bibfnamefont {M.}~\bibnamefont {Tsoi}},\ }\Doi {10.1103/PhysRevX.4.041034}
  {\bibfield  {journal} {\bibinfo  {journal} {Phys. Rev. X},\ }\textbf
  {\bibinfo {volume} {4}},\ \bibinfo {pages} {041034} (\bibinfo {year}
  {2014})}\BibitemShut {NoStop}%
\bibitem [{\citenamefont {N\'u\~nez}\ \emph {et~al.}(2006)\citenamefont
  {N\'u\~nez}, \citenamefont {Duine}, \citenamefont {Haney},\ and\
  \citenamefont {MacDonald}}]{PhysRevB.73.214426}%
  \BibitemOpen
  \bibfield  {author} {\bibinfo {author} {\bibfnamefont {A.~S.}\ \bibnamefont
  {N\'u\~nez}}, \bibinfo {author} {\bibfnamefont {R.~A.}\ \bibnamefont
  {Duine}}, \bibinfo {author} {\bibfnamefont {P.}~\bibnamefont {Haney}}, \ and\
  \bibinfo {author} {\bibfnamefont {A.~H.}\ \bibnamefont {MacDonald}},\ }\href
  {http://link.aps.org/doi/10.1103/PhysRevB.73.214426} {\bibfield  {journal}
  {\bibinfo  {journal} {Phys. Rev. B},\ }\textbf {\bibinfo {volume} {73}},\
  \bibinfo {pages} {214426} (\bibinfo {year} {2006})}\BibitemShut {NoStop}%
\bibitem [{\citenamefont {Swaving}\ and\ \citenamefont
  {Duine}(2011)}]{PhysRevB.83.054428}%
  \BibitemOpen
  \bibfield  {author} {\bibinfo {author} {\bibfnamefont {A.~C.}\ \bibnamefont
  {Swaving}}\ and\ \bibinfo {author} {\bibfnamefont {R.~A.}\ \bibnamefont
  {Duine}},\ }\href {http://link.aps.org/doi/10.1103/PhysRevB.83.054428}
  {\bibfield  {journal} {\bibinfo  {journal} {Phys. Rev. B},\ }\textbf
  {\bibinfo {volume} {83}},\ \bibinfo {pages} {054428} (\bibinfo {year}
  {2011})}\BibitemShut {NoStop}%
\bibitem [{\citenamefont {Linder}(2011)}]{PhysRevB.84.094404}%
  \BibitemOpen
  \bibfield  {author} {\bibinfo {author} {\bibfnamefont {J.}~\bibnamefont
  {Linder}},\ }\Doi {10.1103/PhysRevB.84.094404} {\bibfield  {journal}
  {\bibinfo  {journal} {Phys. Rev. B},\ }\textbf {\bibinfo {volume} {84}},\
  \bibinfo {pages} {094404} (\bibinfo {year} {2011})}\BibitemShut {NoStop}%
\bibitem [{\citenamefont {Gomonay}\ and\ \citenamefont
  {Loktev}(2010)}]{PhysRevB.81.144427}%
  \BibitemOpen
  \bibfield  {author} {\bibinfo {author} {\bibfnamefont {H.~V.}\ \bibnamefont
  {Gomonay}}\ and\ \bibinfo {author} {\bibfnamefont {V.~M.}\ \bibnamefont
  {Loktev}},\ }\href {http://link.aps.org/doi/10.1103/PhysRevB.81.144427}
  {\bibfield  {journal} {\bibinfo  {journal} {Phys. Rev. B},\ }\textbf
  {\bibinfo {volume} {81}},\ \bibinfo {pages} {144427} (\bibinfo {year}
  {2010})}\BibitemShut {NoStop}%
\bibitem [{\citenamefont {Wei}\ \emph {et~al.}(2007)\citenamefont {Wei},
  \citenamefont {Sharma}, \citenamefont {Nunez}, \citenamefont {Haney},
  \citenamefont {Duine}, \citenamefont {Bass}, \citenamefont {MacDonald},\ and\
  \citenamefont {Tsoi}}]{PhysRevLett.98.116603}%
  \BibitemOpen
  \bibfield  {author} {\bibinfo {author} {\bibfnamefont {Z.}~\bibnamefont
  {Wei}}, \bibinfo {author} {\bibfnamefont {A.}~\bibnamefont {Sharma}},
  \bibinfo {author} {\bibfnamefont {A.~S.}\ \bibnamefont {Nunez}}, \bibinfo
  {author} {\bibfnamefont {P.~M.}\ \bibnamefont {Haney}}, \bibinfo {author}
  {\bibfnamefont {R.~A.}\ \bibnamefont {Duine}}, \bibinfo {author}
  {\bibfnamefont {J.}~\bibnamefont {Bass}}, \bibinfo {author} {\bibfnamefont
  {A.~H.}\ \bibnamefont {MacDonald}}, \ and\ \bibinfo {author} {\bibfnamefont
  {M.}~\bibnamefont {Tsoi}},\ }\href
  {http://link.aps.org/doi/10.1103/PhysRevLett.98.116603} {\bibfield  {journal}
  {\bibinfo  {journal} {Phys. Rev. Lett.},\ }\textbf {\bibinfo {volume} {98}},\
  \bibinfo {pages} {116603} (\bibinfo {year} {2007})}\BibitemShut {NoStop}%
\bibitem [{\citenamefont {Cheng}\ \emph {et~al.}(2014)\citenamefont {Cheng},
  \citenamefont {Xiao}, \citenamefont {Niu},\ and\ \citenamefont
  {Brataas}}]{PhysRevLett.113.057601}%
  \BibitemOpen
  \bibfield  {author} {\bibinfo {author} {\bibfnamefont {R.}~\bibnamefont
  {Cheng}}, \bibinfo {author} {\bibfnamefont {J.}~\bibnamefont {Xiao}},
  \bibinfo {author} {\bibfnamefont {Q.}~\bibnamefont {Niu}}, \ and\ \bibinfo
  {author} {\bibfnamefont {A.}~\bibnamefont {Brataas}},\ }\Doi
  {10.1103/PhysRevLett.113.057601} {\bibfield  {journal} {\bibinfo  {journal}
  {Phys. Rev. Lett.},\ }\textbf {\bibinfo {volume} {113}},\ \bibinfo {pages}
  {057601} (\bibinfo {year} {2014})}\BibitemShut {NoStop}%
\bibitem [{\citenamefont {Hals}\ \emph {et~al.}(2011)\citenamefont {Hals},
  \citenamefont {Tserkovnyak},\ and\ \citenamefont
  {Brataas}}]{PhysRevLett.106.107206}%
  \BibitemOpen
  \bibfield  {author} {\bibinfo {author} {\bibfnamefont {K.~M.~D.}\
  \bibnamefont {Hals}}, \bibinfo {author} {\bibfnamefont {Y.}~\bibnamefont
  {Tserkovnyak}}, \ and\ \bibinfo {author} {\bibfnamefont {A.}~\bibnamefont
  {Brataas}},\ }\href {http://link.aps.org/doi/10.1103/PhysRevLett.106.107206}
  {\bibfield  {journal} {\bibinfo  {journal} {Phys. Rev. Lett.},\ }\textbf
  {\bibinfo {volume} {106}},\ \bibinfo {pages} {107206} (\bibinfo {year}
  {2011})}\BibitemShut {NoStop}%
\bibitem [{\citenamefont {Swaving}\ and\ \citenamefont
  {Duine}(2012)}]{JPhysCondMat.24.024223}%
  \BibitemOpen
  \bibfield  {author} {\bibinfo {author} {\bibfnamefont {A.~C.}\ \bibnamefont
  {Swaving}}\ and\ \bibinfo {author} {\bibfnamefont {R.~A.}\ \bibnamefont
  {Duine}},\ }\href {http://stacks.iop.org/0953-8984/24/i=2/a=024223}
  {\bibfield  {journal} {\bibinfo  {journal} {J. Phys.: Cond. Mat.},\ }\textbf
  {\bibinfo {volume} {24}},\ \bibinfo {pages} {024223} (\bibinfo {year}
  {2012})}\BibitemShut {NoStop}%
\bibitem [{\citenamefont {Tveten}\ \emph {et~al.}(2013)\citenamefont {Tveten},
  \citenamefont {Qaiumzadeh}, \citenamefont {Tretiakov},\ and\ \citenamefont
  {Brataas}}]{PhysRevLett.110.127208}%
  \BibitemOpen
  \bibfield  {author} {\bibinfo {author} {\bibfnamefont {E.~G.}\ \bibnamefont
  {Tveten}}, \bibinfo {author} {\bibfnamefont {A.}~\bibnamefont {Qaiumzadeh}},
  \bibinfo {author} {\bibfnamefont {O.~A.}\ \bibnamefont {Tretiakov}}, \ and\
  \bibinfo {author} {\bibfnamefont {A.}~\bibnamefont {Brataas}},\ }\Doi
  {10.1103/PhysRevLett.110.127208} {\bibfield  {journal} {\bibinfo  {journal}
  {Phys. Rev. Lett.},\ }\textbf {\bibinfo {volume} {110}},\ \bibinfo {pages}
  {127208} (\bibinfo {year} {2013})}\BibitemShut {NoStop}%
\bibitem [{\citenamefont {Cheng}\ and\ \citenamefont
  {Niu}(2014)}]{PhysRevB.89.081105}%
  \BibitemOpen
  \bibfield  {author} {\bibinfo {author} {\bibfnamefont {R.}~\bibnamefont
  {Cheng}}\ and\ \bibinfo {author} {\bibfnamefont {Q.}~\bibnamefont {Niu}},\
  }\Doi {10.1103/PhysRevB.89.081105} {\bibfield  {journal} {\bibinfo  {journal}
  {Phys. Rev. B},\ }\textbf {\bibinfo {volume} {89}},\ \bibinfo {pages}
  {081105} (\bibinfo {year} {2014})}\BibitemShut {NoStop}%
\bibitem [{\citenamefont {Tveten}\ \emph {et~al.}(2014)\citenamefont {Tveten},
  \citenamefont {Qaiumzadeh},\ and\ \citenamefont {Brataas}}]{tvetenPRL14}%
  \BibitemOpen
  \bibfield  {author} {\bibinfo {author} {\bibfnamefont {E.~G.}\ \bibnamefont
  {Tveten}}, \bibinfo {author} {\bibfnamefont {A.}~\bibnamefont {Qaiumzadeh}},
  \ and\ \bibinfo {author} {\bibfnamefont {A.}~\bibnamefont {Brataas}},\ }\Doi
  {10.1103/PhysRevLett.112.147204} {\bibfield  {journal} {\bibinfo  {journal}
  {Phys. Rev. Lett.},\ }\textbf {\bibinfo {volume} {112}},\ \bibinfo {pages}
  {147204} (\bibinfo {year} {2014})}\BibitemShut {NoStop}%
\bibitem [{\citenamefont {Kim}\ \emph {et~al.}(2014)\citenamefont {Kim},
  \citenamefont {Tserkovnyak},\ and\ \citenamefont
  {Tchernyshyov}}]{PhysRevB.90.104406}%
  \BibitemOpen
  \bibfield  {author} {\bibinfo {author} {\bibfnamefont {S.~K.}\ \bibnamefont
  {Kim}}, \bibinfo {author} {\bibfnamefont {Y.}~\bibnamefont {Tserkovnyak}}, \
  and\ \bibinfo {author} {\bibfnamefont {O.}~\bibnamefont {Tchernyshyov}},\
  }\Doi {10.1103/PhysRevB.90.104406} {\bibfield  {journal} {\bibinfo  {journal}
  {Phys. Rev. B},\ }\textbf {\bibinfo {volume} {90}},\ \bibinfo {pages}
  {104406} (\bibinfo {year} {2014})}\BibitemShut {NoStop}%
\bibitem [{\citenamefont {Herranz}\ \emph {et~al.}(2009)\citenamefont
  {Herranz}, \citenamefont {Guerrero}, \citenamefont {Villar}, \citenamefont
  {Aliev}, \citenamefont {Swaving}, \citenamefont {Duine}, \citenamefont {van
  Haesendonck},\ and\ \citenamefont {Vavra}}]{PhysRevB.79.134423}%
  \BibitemOpen
  \bibfield  {author} {\bibinfo {author} {\bibfnamefont {D.}~\bibnamefont
  {Herranz}}, \bibinfo {author} {\bibfnamefont {R.}~\bibnamefont {Guerrero}},
  \bibinfo {author} {\bibfnamefont {R.}~\bibnamefont {Villar}}, \bibinfo
  {author} {\bibfnamefont {F.~G.}\ \bibnamefont {Aliev}}, \bibinfo {author}
  {\bibfnamefont {A.~C.}\ \bibnamefont {Swaving}}, \bibinfo {author}
  {\bibfnamefont {R.~A.}\ \bibnamefont {Duine}}, \bibinfo {author}
  {\bibfnamefont {C.}~\bibnamefont {van Haesendonck}}, \ and\ \bibinfo {author}
  {\bibfnamefont {I.}~\bibnamefont {Vavra}},\ }\href
  {http://link.aps.org/doi/10.1103/PhysRevB.79.134423} {\bibfield  {journal}
  {\bibinfo  {journal} {Phys. Rev. B},\ }\textbf {\bibinfo {volume} {79}},\
  \bibinfo {pages} {134423} (\bibinfo {year} {2009})}\BibitemShut {NoStop}%
\bibitem [{\citenamefont {Papanicolaou}(1995)}]{PhysRevB.51.15062}%
  \BibitemOpen
  \bibfield  {author} {\bibinfo {author} {\bibfnamefont {N.}~\bibnamefont
  {Papanicolaou}},\ }\Doi {10.1103/PhysRevB.51.15062} {\bibfield  {journal}
  {\bibinfo  {journal} {Phys. Rev. B},\ }\textbf {\bibinfo {volume} {51}},\
  \bibinfo {pages} {15062} (\bibinfo {year} {1995})}\BibitemShut {NoStop}%
\bibitem [{\citenamefont {Ivanov}\ and\ \citenamefont
  {Kolezhuk}(1995)}]{PhysRevLett.74.1859}%
  \BibitemOpen
  \bibfield  {author} {\bibinfo {author} {\bibfnamefont {B.~A.}\ \bibnamefont
  {Ivanov}}\ and\ \bibinfo {author} {\bibfnamefont {A.~K.}\ \bibnamefont
  {Kolezhuk}},\ }\Doi {10.1103/PhysRevLett.74.1859} {\bibfield  {journal}
  {\bibinfo  {journal} {Phys. Rev. Lett.},\ }\textbf {\bibinfo {volume} {74}},\
  \bibinfo {pages} {1859} (\bibinfo {year} {1995})}\BibitemShut {NoStop}%
\bibitem [{\citenamefont {{Ivanov}}\ and\ \citenamefont
  {{Kolezhuk}}(1995)}]{1995FizNT..21..355I}%
  \BibitemOpen
  \bibfield  {author} {\bibinfo {author} {\bibfnamefont {B.~A.}\ \bibnamefont
  {{Ivanov}}}\ and\ \bibinfo {author} {\bibfnamefont {A.~K.}\ \bibnamefont
  {{Kolezhuk}}},\ }\href@noop {} {\bibfield  {journal} {\bibinfo  {journal}
  {Fiz. Niz. Temp},\ }\textbf {\bibinfo {volume} {21}},\ \bibinfo {pages} {355}
  (\bibinfo {year} {1995})}\BibitemShut {NoStop}%
\bibitem [{\citenamefont {Braun}\ \emph {et~al.}(2005)\citenamefont {Braun},
  \citenamefont {Kulda}, \citenamefont {Roessli}, \citenamefont {Visser},
  \citenamefont {Kramer}, \citenamefont {Gudel},\ and\ \citenamefont
  {Boni}}]{Braun:2005fk}%
  \BibitemOpen
  \bibfield  {author} {\bibinfo {author} {\bibfnamefont {H.-B.}\ \bibnamefont
  {Braun}}, \bibinfo {author} {\bibfnamefont {J.}~\bibnamefont {Kulda}},
  \bibinfo {author} {\bibfnamefont {B.}~\bibnamefont {Roessli}}, \bibinfo
  {author} {\bibfnamefont {D.}~\bibnamefont {Visser}}, \bibinfo {author}
  {\bibfnamefont {K.~W.}\ \bibnamefont {Kramer}}, \bibinfo {author}
  {\bibfnamefont {H.-U.}\ \bibnamefont {Gudel}}, \ and\ \bibinfo {author}
  {\bibfnamefont {P.}~\bibnamefont {Boni}},\ }\href
  {http://dx.doi.org/10.1038/nphys152} {\bibfield  {journal} {\bibinfo
  {journal} {Nat Phys},\ }\textbf {\bibinfo {volume} {1}},\ \bibinfo {pages}
  {159} (\bibinfo {year} {2005})}\BibitemShut {NoStop}%
\bibitem [{\citenamefont {Bar'yakhatar}\ and\ \citenamefont
  {Ivanov}(1979)}]{bar1979nonlinear}%
  \BibitemOpen
  \bibfield  {author} {\bibinfo {author} {\bibfnamefont {I.}~\bibnamefont
  {Bar'yakhatar}}\ and\ \bibinfo {author} {\bibfnamefont {B.}~\bibnamefont
  {Ivanov}},\ }\href@noop {} {\bibfield  {journal} {\bibinfo  {journal} {Fiz.
  Niz. Temp},\ }\textbf {\bibinfo {volume} {5}},\ \bibinfo {pages} {759}
  (\bibinfo {year} {1979})}\BibitemShut {NoStop}%
\bibitem [{\citenamefont {Bar'yakhtar}\ and\ \citenamefont
  {Ivanov}(1980)}]{Baryakhtar:1980qf}%
  \BibitemOpen
  \bibfield  {author} {\bibinfo {author} {\bibfnamefont {I.~V.}\ \bibnamefont
  {Bar'yakhtar}}\ and\ \bibinfo {author} {\bibfnamefont {B.~A.}\ \bibnamefont
  {Ivanov}},\ }\Doi {http://dx.doi.org/10.1016/0038-1098(80)90148-9} {\bibfield
   {journal} {\bibinfo  {journal} {Solid State Communications},\ }\textbf
  {\bibinfo {volume} {34}},\ \bibinfo {pages} {545} (\bibinfo {year}
  {1980})}\BibitemShut {NoStop}%
\bibitem [{\citenamefont {Haldane}(1983)}]{PhysRevLett.50.1153}%
  \BibitemOpen
  \bibfield  {author} {\bibinfo {author} {\bibfnamefont {F.~D.~M.}\
  \bibnamefont {Haldane}},\ }\Doi {10.1103/PhysRevLett.50.1153} {\bibfield
  {journal} {\bibinfo  {journal} {Phys. Rev. Lett.},\ }\textbf {\bibinfo
  {volume} {50}},\ \bibinfo {pages} {1153} (\bibinfo {year}
  {1983})}\BibitemShut {NoStop}%
\bibitem [{\citenamefont {Bar'yakhtar}\ \emph {et~al.}(1985)\citenamefont
  {Bar'yakhtar}, \citenamefont {Ivanov},\ and\ \citenamefont
  {Chetkin}}]{SovPhysUspBaryakhtar1985}%
  \BibitemOpen
  \bibfield  {author} {\bibinfo {author} {\bibfnamefont {V.~G.}\ \bibnamefont
  {Bar'yakhtar}}, \bibinfo {author} {\bibfnamefont {B.~A.}\ \bibnamefont
  {Ivanov}}, \ and\ \bibinfo {author} {\bibfnamefont {M.~V.}\ \bibnamefont
  {Chetkin}},\ }\href {http://stacks.iop.org/0038-5670/28/i=7/a=R02} {\bibfield
   {journal} {\bibinfo  {journal} {Sov. Phys. Usp},\ }\textbf {\bibinfo
  {volume} {28}},\ \bibinfo {pages} {563} (\bibinfo {year} {1985})}\BibitemShut
  {NoStop}%
\bibitem [{\citenamefont {Bode}\ \emph {et~al.}(2006)\citenamefont {Bode},
  \citenamefont {Vedmedenko}, \citenamefont {von Bergmann}, \citenamefont
  {Kubetzka}, \citenamefont {Ferriani}, \citenamefont {Heinze},\ and\
  \citenamefont {Wiesendanger}}]{Bode:2006uq}%
  \BibitemOpen
  \bibfield  {author} {\bibinfo {author} {\bibfnamefont {M.}~\bibnamefont
  {Bode}}, \bibinfo {author} {\bibfnamefont {E.~Y.}\ \bibnamefont
  {Vedmedenko}}, \bibinfo {author} {\bibfnamefont {K.}~\bibnamefont {von
  Bergmann}}, \bibinfo {author} {\bibfnamefont {A.}~\bibnamefont {Kubetzka}},
  \bibinfo {author} {\bibfnamefont {P.}~\bibnamefont {Ferriani}}, \bibinfo
  {author} {\bibfnamefont {S.}~\bibnamefont {Heinze}}, \ and\ \bibinfo {author}
  {\bibfnamefont {R.}~\bibnamefont {Wiesendanger}},\ }\href
  {http://dx.doi.org/10.1038/nmat1646} {\bibfield  {journal} {\bibinfo
  {journal} {Nat. Mater.},\ }\textbf {\bibinfo {volume} {5}},\ \bibinfo {pages}
  {477} (\bibinfo {year} {2006})}\BibitemShut {NoStop}%
\bibitem [{\citenamefont {Jaramillo}\ \emph {et~al.}(2007)\citenamefont
  {Jaramillo}, \citenamefont {Rosenbaum}, \citenamefont {Isaacs}, \citenamefont
  {Shpyrko}, \citenamefont {Evans}, \citenamefont {Aeppli},\ and\ \citenamefont
  {Cai}}]{PhysRevLett.98.117206}%
  \BibitemOpen
  \bibfield  {author} {\bibinfo {author} {\bibfnamefont {R.}~\bibnamefont
  {Jaramillo}}, \bibinfo {author} {\bibfnamefont {T.~F.}\ \bibnamefont
  {Rosenbaum}}, \bibinfo {author} {\bibfnamefont {E.~D.}\ \bibnamefont
  {Isaacs}}, \bibinfo {author} {\bibfnamefont {O.~G.}\ \bibnamefont {Shpyrko}},
  \bibinfo {author} {\bibfnamefont {P.~G.}\ \bibnamefont {Evans}}, \bibinfo
  {author} {\bibfnamefont {G.}~\bibnamefont {Aeppli}}, \ and\ \bibinfo {author}
  {\bibfnamefont {Z.}~\bibnamefont {Cai}},\ }\href
  {http://link.aps.org/doi/10.1103/PhysRevLett.98.117206} {\bibfield  {journal}
  {\bibinfo  {journal} {Phys. Rev. Lett.},\ }\textbf {\bibinfo {volume} {98}},\
  \bibinfo {pages} {117206} (\bibinfo {year} {2007})}\BibitemShut {NoStop}%
\bibitem [{\citenamefont {Weber}\ \emph {et~al.}(2003)\citenamefont {Weber},
  \citenamefont {Ohldag}, \citenamefont {Gomonaj},\ and\ \citenamefont
  {Hillebrecht}}]{PhysRevLett.91.237205}%
  \BibitemOpen
  \bibfield  {author} {\bibinfo {author} {\bibfnamefont {N.~B.}\ \bibnamefont
  {Weber}}, \bibinfo {author} {\bibfnamefont {H.}~\bibnamefont {Ohldag}},
  \bibinfo {author} {\bibfnamefont {H.}~\bibnamefont {Gomonaj}}, \ and\
  \bibinfo {author} {\bibfnamefont {F.~U.}\ \bibnamefont {Hillebrecht}},\ }\Doi
  {10.1103/PhysRevLett.91.237205} {\bibfield  {journal} {\bibinfo  {journal}
  {Phys. Rev. Lett.},\ }\textbf {\bibinfo {volume} {91}},\ \bibinfo {pages}
  {237205} (\bibinfo {year} {2003})}\BibitemShut {NoStop}%
\bibitem [{\citenamefont {Bezencenet}\ \emph {et~al.}(2011)\citenamefont
  {Bezencenet}, \citenamefont {Bonamy}, \citenamefont {Belkhou}, \citenamefont
  {Ohresser},\ and\ \citenamefont {Barbier}}]{PhysRevLett.106.107201}%
  \BibitemOpen
  \bibfield  {author} {\bibinfo {author} {\bibfnamefont {O.}~\bibnamefont
  {Bezencenet}}, \bibinfo {author} {\bibfnamefont {D.}~\bibnamefont {Bonamy}},
  \bibinfo {author} {\bibfnamefont {R.}~\bibnamefont {Belkhou}}, \bibinfo
  {author} {\bibfnamefont {P.}~\bibnamefont {Ohresser}}, \ and\ \bibinfo
  {author} {\bibfnamefont {A.}~\bibnamefont {Barbier}},\ }\Doi
  {10.1103/PhysRevLett.106.107201} {\bibfield  {journal} {\bibinfo  {journal}
  {Phys. Rev. Lett.},\ }\textbf {\bibinfo {volume} {106}},\ \bibinfo {pages}
  {107201} (\bibinfo {year} {2011})}\BibitemShut {NoStop}%
\bibitem [{\citenamefont {Czekaj}\ \emph {et~al.}(2006)\citenamefont {Czekaj},
  \citenamefont {Nolting}, \citenamefont {Heyderman}, \citenamefont
  {Willmott},\ and\ \citenamefont {van~der Laan}}]{PhysRevB.73.020401}%
  \BibitemOpen
  \bibfield  {author} {\bibinfo {author} {\bibfnamefont {S.}~\bibnamefont
  {Czekaj}}, \bibinfo {author} {\bibfnamefont {F.}~\bibnamefont {Nolting}},
  \bibinfo {author} {\bibfnamefont {L.~J.}\ \bibnamefont {Heyderman}}, \bibinfo
  {author} {\bibfnamefont {P.~R.}\ \bibnamefont {Willmott}}, \ and\ \bibinfo
  {author} {\bibfnamefont {G.}~\bibnamefont {van~der Laan}},\ }\Doi
  {10.1103/PhysRevB.73.020401} {\bibfield  {journal} {\bibinfo  {journal}
  {Phys. Rev. B},\ }\textbf {\bibinfo {volume} {73}},\ \bibinfo {pages}
  {020401} (\bibinfo {year} {2006})}\BibitemShut {NoStop}%
\bibitem [{\citenamefont {Folven}\ \emph {et~al.}(2010)\citenamefont {Folven},
  \citenamefont {Tybell}, \citenamefont {Scholl}, \citenamefont {Young},
  \citenamefont {Retterer}, \citenamefont {Takamura},\ and\ \citenamefont
  {Grepstad}}]{doi:10.1021/nl1025908}%
  \BibitemOpen
  \bibfield  {author} {\bibinfo {author} {\bibfnamefont {E.}~\bibnamefont
  {Folven}}, \bibinfo {author} {\bibfnamefont {T.}~\bibnamefont {Tybell}},
  \bibinfo {author} {\bibfnamefont {A.}~\bibnamefont {Scholl}}, \bibinfo
  {author} {\bibfnamefont {A.}~\bibnamefont {Young}}, \bibinfo {author}
  {\bibfnamefont {S.~T.}\ \bibnamefont {Retterer}}, \bibinfo {author}
  {\bibfnamefont {Y.}~\bibnamefont {Takamura}}, \ and\ \bibinfo {author}
  {\bibfnamefont {J.~K.}\ \bibnamefont {Grepstad}},\ }\Doi {10.1021/nl1025908}
  {\bibfield  {journal} {\bibinfo  {journal} {Nano Letters},\ }\textbf
  {\bibinfo {volume} {10}},\ \bibinfo {pages} {4578} (\bibinfo {year}
  {2010})},\ \bibinfo {note} {pMID: 20942384},\ \Eprint
  {http://arxiv.org/abs/http://dx.doi.org/10.1021/nl1025908}
  {http://dx.doi.org/10.1021/nl1025908} \BibitemShut {NoStop}%
\bibitem [{\citenamefont {Cabra}\ and\ \citenamefont
  {Pujol}(2004)}]{CabraPujol2004}%
  \BibitemOpen
  \bibfield  {author} {\bibinfo {author} {\bibfnamefont {D.}~\bibnamefont
  {Cabra}}\ and\ \bibinfo {author} {\bibfnamefont {P.}~\bibnamefont {Pujol}},\
  }in\ \Doi {10.1007/BFb0119596} {\emph {\bibinfo {booktitle} {Quantum
  Magnetism}}},\ \bibinfo {series} {Lecture Notes in Physics}, Vol.\ \bibinfo
  {volume} {645},\ \bibinfo {editor} {edited by\ \bibinfo {editor}
  {\bibfnamefont {U.}~\bibnamefont {Schollw{\"o}ck}}, \bibinfo {editor}
  {\bibfnamefont {J.}~\bibnamefont {Richter}}, \bibinfo {editor} {\bibfnamefont
  {D.}~\bibnamefont {Farnell}}, \ and\ \bibinfo {editor} {\bibfnamefont
  {R.}~\bibnamefont {Bishop}}}\ (\bibinfo  {publisher} {Springer Berlin
  Heidelberg},\ \bibinfo {year} {2004})\ pp.\ \bibinfo {pages} {253--305},\
  ISBN \bibinfo {isbn} {978-3-540-21422-9}\BibitemShut {NoStop}%
\bibitem [{\citenamefont {Auerbach}(2012)}]{auerbach2012interacting}%
  \BibitemOpen
  \bibfield  {author} {\bibinfo {author} {\bibfnamefont {A.}~\bibnamefont
  {Auerbach}},\ }\href@noop {} {\emph {\bibinfo {title} {Interacting electrons
  and quantum magnetism}}}\ (\bibinfo  {publisher} {Springer Science \&
  Business Media},\ \bibinfo {year} {2012})\BibitemShut {NoStop}%
\bibitem [{\citenamefont {Affleck}(1985)}]{AffleckNPB1985}%
  \BibitemOpen
  \bibfield  {author} {\bibinfo {author} {\bibfnamefont {I.}~\bibnamefont
  {Affleck}},\ }\Doi {http://dx.doi.org/10.1016/0550-3213(85)90353-0}
  {\bibfield  {journal} {\bibinfo  {journal} {Nuclear Physics B},\ }\textbf
  {\bibinfo {volume} {257}},\ \bibinfo {pages} {397} (\bibinfo {year}
  {1985})}\BibitemShut {NoStop}%
\bibitem [{\citenamefont {Haldane}(1985)}]{Haldane:1985sf}%
  \BibitemOpen
  \bibfield  {author} {\bibinfo {author} {\bibfnamefont {F.~D.~M.}\
  \bibnamefont {Haldane}},\ }\href
  {http://scitation.aip.org/content/aip/journal/jap/57/8/10.1063/1.335096}
  {\bibfield  {journal} {\bibinfo  {journal} {J. Appl. Phys.},\ }\textbf
  {\bibinfo {volume} {57}},\ \bibinfo {pages} {3359} (\bibinfo {year}
  {1985})}\BibitemShut {NoStop}%
\bibitem [{\citenamefont {Ivanov}\ and\ \citenamefont
  {Kolezhuk}(1997)}]{PhysRevB.56.8886}%
  \BibitemOpen
  \bibfield  {author} {\bibinfo {author} {\bibfnamefont {B.~A.}\ \bibnamefont
  {Ivanov}}\ and\ \bibinfo {author} {\bibfnamefont {A.~K.}\ \bibnamefont
  {Kolezhuk}},\ }\Doi {10.1103/PhysRevB.56.8886} {\bibfield  {journal}
  {\bibinfo  {journal} {Phys. Rev. B},\ }\textbf {\bibinfo {volume} {56}},\
  \bibinfo {pages} {8886} (\bibinfo {year} {1997})}\BibitemShut {NoStop}%
\bibitem [{\citenamefont {Papanicolaou}(1998)}]{PapanicolaouJPCM1998}%
  \BibitemOpen
  \bibfield  {author} {\bibinfo {author} {\bibfnamefont {N.}~\bibnamefont
  {Papanicolaou}},\ }\href {http://stacks.iop.org/0953-8984/10/i=8/a=004}
  {\bibfield  {journal} {\bibinfo  {journal} {J. Phys.: Cond. Mat.},\ }\textbf
  {\bibinfo {volume} {10}} (\bibinfo {year} {1998})}\BibitemShut {NoStop}%
\bibitem [{\citenamefont {Anderson}(1952)}]{PhysRev.86.694}%
  \BibitemOpen
  \bibfield  {author} {\bibinfo {author} {\bibfnamefont {P.~W.}\ \bibnamefont
  {Anderson}},\ }\Doi {10.1103/PhysRev.86.694} {\bibfield  {journal} {\bibinfo
  {journal} {Phys. Rev.},\ }\textbf {\bibinfo {volume} {86}},\ \bibinfo {pages}
  {694} (\bibinfo {year} {1952})}\BibitemShut {NoStop}%
\bibitem [{\citenamefont {Affleck}(1989)}]{0953-8984-1-19-001}%
  \BibitemOpen
  \bibfield  {author} {\bibinfo {author} {\bibfnamefont {I.}~\bibnamefont
  {Affleck}},\ }\href {http://stacks.iop.org/0953-8984/1/i=19/a=001} {\bibfield
   {journal} {\bibinfo  {journal} {Journal of Physics: Condensed Matter},\
  }\textbf {\bibinfo {volume} {1}},\ \bibinfo {pages} {3047} (\bibinfo {year}
  {1989})}\BibitemShut {NoStop}%
\bibitem [{\citenamefont {Lifshitz}\ and\ \citenamefont
  {Pitaevskii}(1980)}]{lifshitz1980}%
  \BibitemOpen
  \bibfield  {author} {\bibinfo {author} {\bibfnamefont {E.~M.}\ \bibnamefont
  {Lifshitz}}\ and\ \bibinfo {author} {\bibfnamefont {L.~P.}\ \bibnamefont
  {Pitaevskii}},\ }\href@noop {} {\emph {\bibinfo {title} {Statistical Physics,
  Course of Theoretical Physics}}},\ Vol.~\bibinfo {volume} {9}\ (\bibinfo
  {publisher} {Pergamon, Oxford},\ \bibinfo {year} {1980})\BibitemShut
  {NoStop}%
\bibitem [{\citenamefont {Braun}\ and\ \citenamefont
  {Loss}(1996)}]{PhysRevB.53.3237}%
  \BibitemOpen
  \bibfield  {author} {\bibinfo {author} {\bibfnamefont {H.-B.}\ \bibnamefont
  {Braun}}\ and\ \bibinfo {author} {\bibfnamefont {D.}~\bibnamefont {Loss}},\
  }\Doi {10.1103/PhysRevB.53.3237} {\bibfield  {journal} {\bibinfo  {journal}
  {Phys. Rev. B},\ }\textbf {\bibinfo {volume} {53}},\ \bibinfo {pages} {3237}
  (\bibinfo {year} {1996})}\BibitemShut {NoStop}%
\bibitem [{\citenamefont {Tatara}\ \emph {et~al.}(2008)\citenamefont {Tatara},
  \citenamefont {Kohno},\ and\ \citenamefont {Shibata}}]{Tatara2008213}%
  \BibitemOpen
  \bibfield  {author} {\bibinfo {author} {\bibfnamefont {G.}~\bibnamefont
  {Tatara}}, \bibinfo {author} {\bibfnamefont {H.}~\bibnamefont {Kohno}}, \
  and\ \bibinfo {author} {\bibfnamefont {J.}~\bibnamefont {Shibata}},\ }\href
  {http://www.sciencedirect.com/science/article/pii/S0370157308002597}
  {\bibfield  {journal} {\bibinfo  {journal} {Phys. Rep.},\ }\textbf {\bibinfo
  {volume} {468}},\ \bibinfo {pages} {213 } (\bibinfo {year}
  {2008})}\BibitemShut {NoStop}%
\bibitem [{\citenamefont {Fradkin}\ and\ \citenamefont
  {Stone}(1988)}]{PhysRevB.38.7215}%
  \BibitemOpen
  \bibfield  {author} {\bibinfo {author} {\bibfnamefont {E.}~\bibnamefont
  {Fradkin}}\ and\ \bibinfo {author} {\bibfnamefont {M.}~\bibnamefont
  {Stone}},\ }\Doi {10.1103/PhysRevB.38.7215} {\bibfield  {journal} {\bibinfo
  {journal} {Phys. Rev. B},\ }\textbf {\bibinfo {volume} {38}},\ \bibinfo
  {pages} {7215} (\bibinfo {year} {1988})}\BibitemShut {NoStop}%
\bibitem [{\citenamefont {Haldane}(1988)}]{PhysRevLett.61.1029}%
  \BibitemOpen
  \bibfield  {author} {\bibinfo {author} {\bibfnamefont {F.~D.~M.}\
  \bibnamefont {Haldane}},\ }\Doi {10.1103/PhysRevLett.61.1029} {\bibfield
  {journal} {\bibinfo  {journal} {Phys. Rev. Lett.},\ }\textbf {\bibinfo
  {volume} {61}},\ \bibinfo {pages} {1029} (\bibinfo {year}
  {1988})}\BibitemShut {NoStop}%
\bibitem [{\citenamefont {Read}\ and\ \citenamefont
  {Sachdev}(1989)}]{Read1989609}%
  \BibitemOpen
  \bibfield  {author} {\bibinfo {author} {\bibfnamefont {N.}~\bibnamefont
  {Read}}\ and\ \bibinfo {author} {\bibfnamefont {S.}~\bibnamefont {Sachdev}},\
  }\Doi {http://dx.doi.org/10.1016/0550-3213(89)90061-8} {\bibfield  {journal}
  {\bibinfo  {journal} {Nuclear Physics B},\ }\textbf {\bibinfo {volume}
  {316}},\ \bibinfo {pages} {609 } (\bibinfo {year} {1989})}\BibitemShut
  {NoStop}%
\bibitem [{\citenamefont {Read}\ and\ \citenamefont
  {Sachdev}(1990)}]{PhysRevB.42.4568}%
  \BibitemOpen
  \bibfield  {author} {\bibinfo {author} {\bibfnamefont {N.}~\bibnamefont
  {Read}}\ and\ \bibinfo {author} {\bibfnamefont {S.}~\bibnamefont {Sachdev}},\
  }\Doi {10.1103/PhysRevB.42.4568} {\bibfield  {journal} {\bibinfo  {journal}
  {Phys. Rev. B},\ }\textbf {\bibinfo {volume} {42}},\ \bibinfo {pages} {4568}
  (\bibinfo {year} {1990})}\BibitemShut {NoStop}%
\bibitem [{\citenamefont {Andreev}\ and\ \citenamefont
  {Marchenko}(1980)}]{SovPhysUsp23.21}%
  \BibitemOpen
  \bibfield  {author} {\bibinfo {author} {\bibfnamefont {A.}~\bibnamefont
  {Andreev}}\ and\ \bibinfo {author} {\bibfnamefont {V.~I.}\ \bibnamefont
  {Marchenko}},\ }\href@noop {} {\bibfield  {journal} {\bibinfo  {journal}
  {Sov. Phys. Usp},\ }\textbf {\bibinfo {volume} {23}} (\bibinfo {year}
  {1980})}\BibitemShut {NoStop}%
\bibitem [{\citenamefont {Oguchi}(1960)}]{PhysRev.117.117}%
  \BibitemOpen
  \bibfield  {author} {\bibinfo {author} {\bibfnamefont {T.}~\bibnamefont
  {Oguchi}},\ }\Doi {10.1103/PhysRev.117.117} {\bibfield  {journal} {\bibinfo
  {journal} {Phys. Rev.},\ }\textbf {\bibinfo {volume} {117}},\ \bibinfo
  {pages} {117} (\bibinfo {year} {1960})}\BibitemShut {NoStop}%
\bibitem [{\citenamefont {Takahashi}(1989)}]{PhysRevB.40.2494}%
  \BibitemOpen
  \bibfield  {author} {\bibinfo {author} {\bibfnamefont {M.}~\bibnamefont
  {Takahashi}},\ }\Doi {10.1103/PhysRevB.40.2494} {\bibfield  {journal}
  {\bibinfo  {journal} {Phys. Rev. B},\ }\textbf {\bibinfo {volume} {40}},\
  \bibinfo {pages} {2494} (\bibinfo {year} {1989})}\BibitemShut {NoStop}%
\bibitem [{\citenamefont {Brataas}\ \emph {et~al.}(2012)\citenamefont
  {Brataas}, \citenamefont {Kent},\ and\ \citenamefont
  {Ohno}}]{Brataas:2012fk}%
  \BibitemOpen
  \bibfield  {author} {\bibinfo {author} {\bibfnamefont {A.}~\bibnamefont
  {Brataas}}, \bibinfo {author} {\bibfnamefont {A.~D.}\ \bibnamefont {Kent}}, \
  and\ \bibinfo {author} {\bibfnamefont {H.}~\bibnamefont {Ohno}},\ }\href
  {http://dx.doi.org/10.1038/nmat3311} {\bibfield  {journal} {\bibinfo
  {journal} {Nat. Mater.},\ }\textbf {\bibinfo {volume} {11}},\ \bibinfo
  {pages} {372} (\bibinfo {year} {2012})}\BibitemShut {NoStop}%
\bibitem [{\citenamefont {Tretiakov}\ \emph {et~al.}(2008)\citenamefont
  {Tretiakov}, \citenamefont {Clarke}, \citenamefont {Chern}, \citenamefont
  {Bazaliy},\ and\ \citenamefont {Tchernyshyov}}]{PhysRevLett.100.127204}%
  \BibitemOpen
  \bibfield  {author} {\bibinfo {author} {\bibfnamefont {O.~A.}\ \bibnamefont
  {Tretiakov}}, \bibinfo {author} {\bibfnamefont {D.}~\bibnamefont {Clarke}},
  \bibinfo {author} {\bibfnamefont {G.-W.}\ \bibnamefont {Chern}}, \bibinfo
  {author} {\bibfnamefont {Y.~B.}\ \bibnamefont {Bazaliy}}, \ and\ \bibinfo
  {author} {\bibfnamefont {O.}~\bibnamefont {Tchernyshyov}},\ }\href
  {http://link.aps.org/doi/10.1103/PhysRevLett.100.127204} {\bibfield
  {journal} {\bibinfo  {journal} {Phys. Rev. Lett.},\ }\textbf {\bibinfo
  {volume} {100}},\ \bibinfo {pages} {127204} (\bibinfo {year}
  {2008})}\BibitemShut {NoStop}%
\bibitem [{\citenamefont {Schryer}\ and\ \citenamefont
  {Walker}(1974)}]{schryer:5406}%
  \BibitemOpen
  \bibfield  {author} {\bibinfo {author} {\bibfnamefont {N.~L.}\ \bibnamefont
  {Schryer}}\ and\ \bibinfo {author} {\bibfnamefont {L.~R.}\ \bibnamefont
  {Walker}},\ }\href {http://link.aip.org/link/?JAP/45/5406/1} {\bibfield
  {journal} {\bibinfo  {journal} {J. Appl. Phys.},\ }\textbf {\bibinfo {volume}
  {45}},\ \bibinfo {pages} {5406} (\bibinfo {year} {1974})}\BibitemShut
  {NoStop}%
\bibitem [{\citenamefont {Kubetzka}\ \emph {et~al.}(2005)\citenamefont
  {Kubetzka}, \citenamefont {Ferriani}, \citenamefont {Bode}, \citenamefont
  {Heinze}, \citenamefont {Bihlmayer}, \citenamefont {von Bergmann},
  \citenamefont {Pietzsch}, \citenamefont {Bl\"ugel},\ and\ \citenamefont
  {Wiesendanger}}]{PhysRevLett.94.087204}%
  \BibitemOpen
  \bibfield  {author} {\bibinfo {author} {\bibfnamefont {A.}~\bibnamefont
  {Kubetzka}}, \bibinfo {author} {\bibfnamefont {P.}~\bibnamefont {Ferriani}},
  \bibinfo {author} {\bibfnamefont {M.}~\bibnamefont {Bode}}, \bibinfo {author}
  {\bibfnamefont {S.}~\bibnamefont {Heinze}}, \bibinfo {author} {\bibfnamefont
  {G.}~\bibnamefont {Bihlmayer}}, \bibinfo {author} {\bibfnamefont
  {K.}~\bibnamefont {von Bergmann}}, \bibinfo {author} {\bibfnamefont
  {O.}~\bibnamefont {Pietzsch}}, \bibinfo {author} {\bibfnamefont
  {S.}~\bibnamefont {Bl\"ugel}}, \ and\ \bibinfo {author} {\bibfnamefont
  {R.}~\bibnamefont {Wiesendanger}},\ }\Doi {10.1103/PhysRevLett.94.087204}
  {\bibfield  {journal} {\bibinfo  {journal} {Phys. Rev. Lett.},\ }\textbf
  {\bibinfo {volume} {94}},\ \bibinfo {pages} {087204} (\bibinfo {year}
  {2005})}\BibitemShut {NoStop}%
\bibitem [{\citenamefont {Wu}\ \emph {et~al.}(2011)\citenamefont {Wu},
  \citenamefont {Carlton}, \citenamefont {Park}, \citenamefont {Meng},
  \citenamefont {Arenholz}, \citenamefont {Doran}, \citenamefont {Young},
  \citenamefont {Scholl}, \citenamefont {Hwang}, \citenamefont {Zhao},
  \citenamefont {Bokor},\ and\ \citenamefont {Qiu}}]{Wu:2011yu}%
  \BibitemOpen
  \bibfield  {author} {\bibinfo {author} {\bibfnamefont {J.}~\bibnamefont
  {Wu}}, \bibinfo {author} {\bibfnamefont {D.}~\bibnamefont {Carlton}},
  \bibinfo {author} {\bibfnamefont {J.~S.}\ \bibnamefont {Park}}, \bibinfo
  {author} {\bibfnamefont {Y.}~\bibnamefont {Meng}}, \bibinfo {author}
  {\bibfnamefont {E.}~\bibnamefont {Arenholz}}, \bibinfo {author}
  {\bibfnamefont {A.}~\bibnamefont {Doran}}, \bibinfo {author} {\bibfnamefont
  {A.~T.}\ \bibnamefont {Young}}, \bibinfo {author} {\bibfnamefont
  {A.}~\bibnamefont {Scholl}}, \bibinfo {author} {\bibfnamefont
  {C.}~\bibnamefont {Hwang}}, \bibinfo {author} {\bibfnamefont {H.~W.}\
  \bibnamefont {Zhao}}, \bibinfo {author} {\bibfnamefont {J.}~\bibnamefont
  {Bokor}}, \ and\ \bibinfo {author} {\bibfnamefont {Z.~Q.}\ \bibnamefont
  {Qiu}},\ }\href {http://dx.doi.org/10.1038/nphys1891} {\bibfield  {journal}
  {\bibinfo  {journal} {Nat. Phys.},\ }\textbf {\bibinfo {volume} {7}},\
  \bibinfo {pages} {303} (\bibinfo {year} {2011})}\BibitemShut {NoStop}%
\bibitem [{\citenamefont {Michel}\ \emph {et~al.}(1991)\citenamefont {Michel},
  \citenamefont {Israeloff}, \citenamefont {Weissman}, \citenamefont {Dura},\
  and\ \citenamefont {Flynn}}]{PhysRevB.44.7413}%
  \BibitemOpen
  \bibfield  {author} {\bibinfo {author} {\bibfnamefont {R.~P.}\ \bibnamefont
  {Michel}}, \bibinfo {author} {\bibfnamefont {N.~E.}\ \bibnamefont
  {Israeloff}}, \bibinfo {author} {\bibfnamefont {M.~B.}\ \bibnamefont
  {Weissman}}, \bibinfo {author} {\bibfnamefont {J.~A.}\ \bibnamefont {Dura}},
  \ and\ \bibinfo {author} {\bibfnamefont {C.~P.}\ \bibnamefont {Flynn}},\
  }\Doi {10.1103/PhysRevB.44.7413} {\bibfield  {journal} {\bibinfo  {journal}
  {Phys. Rev. B},\ }\textbf {\bibinfo {volume} {44}},\ \bibinfo {pages} {7413}
  (\bibinfo {year} {1991})}\BibitemShut {NoStop}%
\bibitem [{\citenamefont {Shpyrko}\ \emph {et~al.}(2007)\citenamefont
  {Shpyrko}, \citenamefont {Isaacs}, \citenamefont {Logan}, \citenamefont
  {Feng}, \citenamefont {Aeppli}, \citenamefont {Jaramillo}, \citenamefont
  {Kim}, \citenamefont {Rosenbaum}, \citenamefont {Zschack}, \citenamefont
  {Sprung}, \citenamefont {Narayanan},\ and\ \citenamefont
  {Sandy}}]{Shpyrko:2007fk}%
  \BibitemOpen
  \bibfield  {author} {\bibinfo {author} {\bibfnamefont {O.~G.}\ \bibnamefont
  {Shpyrko}}, \bibinfo {author} {\bibfnamefont {E.~D.}\ \bibnamefont {Isaacs}},
  \bibinfo {author} {\bibfnamefont {J.~M.}\ \bibnamefont {Logan}}, \bibinfo
  {author} {\bibfnamefont {Y.}~\bibnamefont {Feng}}, \bibinfo {author}
  {\bibfnamefont {G.}~\bibnamefont {Aeppli}}, \bibinfo {author} {\bibfnamefont
  {R.}~\bibnamefont {Jaramillo}}, \bibinfo {author} {\bibfnamefont {H.~C.}\
  \bibnamefont {Kim}}, \bibinfo {author} {\bibfnamefont {T.~F.}\ \bibnamefont
  {Rosenbaum}}, \bibinfo {author} {\bibfnamefont {P.}~\bibnamefont {Zschack}},
  \bibinfo {author} {\bibfnamefont {M.}~\bibnamefont {Sprung}}, \bibinfo
  {author} {\bibfnamefont {S.}~\bibnamefont {Narayanan}}, \ and\ \bibinfo
  {author} {\bibfnamefont {A.~R.}\ \bibnamefont {Sandy}},\ }\href
  {http://dx.doi.org/10.1038/nature05776} {\bibfield  {journal} {\bibinfo
  {journal} {Nature},\ }\textbf {\bibinfo {volume} {447}},\ \bibinfo {pages}
  {68} (\bibinfo {year} {2007})}\BibitemShut {NoStop}%
\bibitem [{\citenamefont {Logan}\ \emph {et~al.}(2012)\citenamefont {Logan},
  \citenamefont {Kim}, \citenamefont {Rosenmann}, \citenamefont {Cai},
  \citenamefont {Divan}, \citenamefont {Shpyrko},\ and\ \citenamefont
  {Isaacs}}]{logan:192405}%
  \BibitemOpen
  \bibfield  {author} {\bibinfo {author} {\bibfnamefont {J.~M.}\ \bibnamefont
  {Logan}}, \bibinfo {author} {\bibfnamefont {H.~C.}\ \bibnamefont {Kim}},
  \bibinfo {author} {\bibfnamefont {D.}~\bibnamefont {Rosenmann}}, \bibinfo
  {author} {\bibfnamefont {Z.}~\bibnamefont {Cai}}, \bibinfo {author}
  {\bibfnamefont {R.}~\bibnamefont {Divan}}, \bibinfo {author} {\bibfnamefont
  {O.~G.}\ \bibnamefont {Shpyrko}}, \ and\ \bibinfo {author} {\bibfnamefont
  {E.~D.}\ \bibnamefont {Isaacs}},\ }\href
  {http://link.aip.org/link/?APL/100/192405/1} {\bibfield  {journal} {\bibinfo
  {journal} {Appl. Phys. Lett.},\ }\textbf {\bibinfo {volume} {100}},\ \bibinfo
  {eid} {192405} (\bibinfo {year} {2012})}\BibitemShut {NoStop}%
\end{thebibliography}
\end{document}